\documentclass[sigconf]{acmart}

\usepackage{multirow}
\usepackage{graphicx,balance}
\usepackage{booktabs}

\usepackage{enumitem}

\usepackage{subfigure}
\usepackage{pgfplots}
\usepackage{tikz}
\usepackage{tikz-qtree}

\usetikzlibrary{matrix}
\usetikzlibrary{patterns}
\usepgfplotslibrary{groupplots}

\definecolor{1}{RGB}{102,102,255}
\definecolor{2}{RGB}{102,206,245}
\definecolor{3}{RGB}{255,247,102}
\definecolor{4}{RGB}{255,178,102}
\definecolor{11}{RGB}{146,209,79}
\definecolor{5}{RGB}{194,209,163}
\definecolor{6}{RGB}{102,163,255}
\definecolor{7}{RGB}{163,223,196}
\definecolor{8}{RGB}{194,163,163}
\definecolor{9}{RGB}{194,102,163}
\definecolor{10}{RGB}{255,102,102}
\definecolor{red1}{RGB}{199,21,133}
\definecolor{red2}{RGB}{255,102,102}
\definecolor{red3}{RGB}{255,182,193}
\definecolor{blue1}{RGB}{25,25,112}
\definecolor{blue2}{RGB}{30,144,255}
\definecolor{blue3}{RGB}{135,206,250}

\usepackage{color}

\AtBeginDocument{%
  \providecommand\BibTeX{{%
    \normalfont B\kern-0.5em{\scshape i\kern-0.25em b}\kern-0.8em\TeX}}}

\copyrightyear{2023}
\acmYear{2023}
\setcopyright{acmlicensed}
\acmConference[RecSys '23]{Seventeenth ACM Conference on Recommender Systems}{September 18--22, 2023}{Singapore, Singapore}
\acmBooktitle{Seventeenth ACM Conference on Recommender Systems (RecSys '23), September 18--22, 2023, Singapore, Singapore}
\acmPrice{15.00}
\acmDOI{10.1145/3604915.3608802}
\acmISBN{979-8-4007-0241-9/23/09}

\begin{document}

\title[Domain Disentanglement with Interpolative Data Augmentation for Dual-Target CDR]{Domain Disentanglement with Interpolative Data Augmentation for Dual-Target Cross-Domain Recommendation}

\author{Jiajie Zhu}
\email{jiajie.zhu1@students.mq.edu.au}
\orcid{0000-0001-8673-1477}
\affiliation{%
  \institution{Macquarie University}
  \country{Australia}
}

\author{Yan Wang}
\authornote{Corresponding author.}
\email{yan.wang@mq.edu.au}
\orcid{0000-0002-5344-1884}
\affiliation{%
  \institution{Macquarie University}
  \country{Australia}
}

\author{Feng Zhu}
\email{zhufeng.zhu@antgroup.com}
\orcid{0000-0003-4200-0423}
\affiliation{%
  \institution{Ant Group}
  \country{China}}

\author{Zhu Sun}
\email{sunzhuntu@gmail.com}
\orcid{0000-0002-3350-7022}
\affiliation{
  \institution{Institute of High Performance Computing and Centre for Frontier AI Research, A*STAR}
  \country{Singapore}}

\begin{abstract}
  The conventional single-target Cross-Domain Recommendation (CDR) aims to improve the recommendation performance on a sparser target domain by transferring the knowledge from a source domain that contains relatively richer information. By contrast, in recent years, dual-target CDR has been proposed to improve the recommendation performance on both domains simultaneously. However, to this end, there are two challenges in dual-target CDR: (1) how to generate both relevant and diverse augmented user representations, and (2) how to effectively decouple domain-independent information from domain-specific information, in addition to domain-shared information, to capture comprehensive user preferences. To address the above two challenges, we propose a \textbf{D}isentanglement-based framework with \textbf{I}nterpolative \textbf{D}ata \textbf{A}ugmentation for dual-target \textbf{C}ross-\textbf{D}omain \textbf{R}ecommendation, called DIDA-CDR. In DIDA-CDR, we first propose an interpolative data augmentation approach to generating both relevant and diverse augmented user representations to augment sparser domain and explore potential user preferences. We then propose a disentanglement module to effectively decouple domain-specific and domain-independent information to capture comprehensive user preferences. Both steps significantly contribute to capturing more comprehensive user preferences, thereby improving the recommendation performance on each domain. Extensive experiments conducted on five real-world datasets show the significant superiority of DIDA-CDR over the state-of-the-art methods.
\end{abstract}




\maketitle

\section{Introduction}
To alleviate the data sparsity problem, Cross-Domain Recommendation (CDR) \cite{loni2014cross} aims to employ abundant information from a relatively richer domain to improve recommendation performance on a sparser domain, forming the so-called \emph{single-target CDR} \cite{zhu2019dtcdr}. By contrast, in recent years, \emph{dual-target CDR} has been proposed to improve the recommendation performance on both domains simultaneously by sharing the common knowledge across domains \cite{ijcai2021p639,zang2021survey}.

\begin{figure*}[ht]
\centering
\includegraphics[scale=0.43]{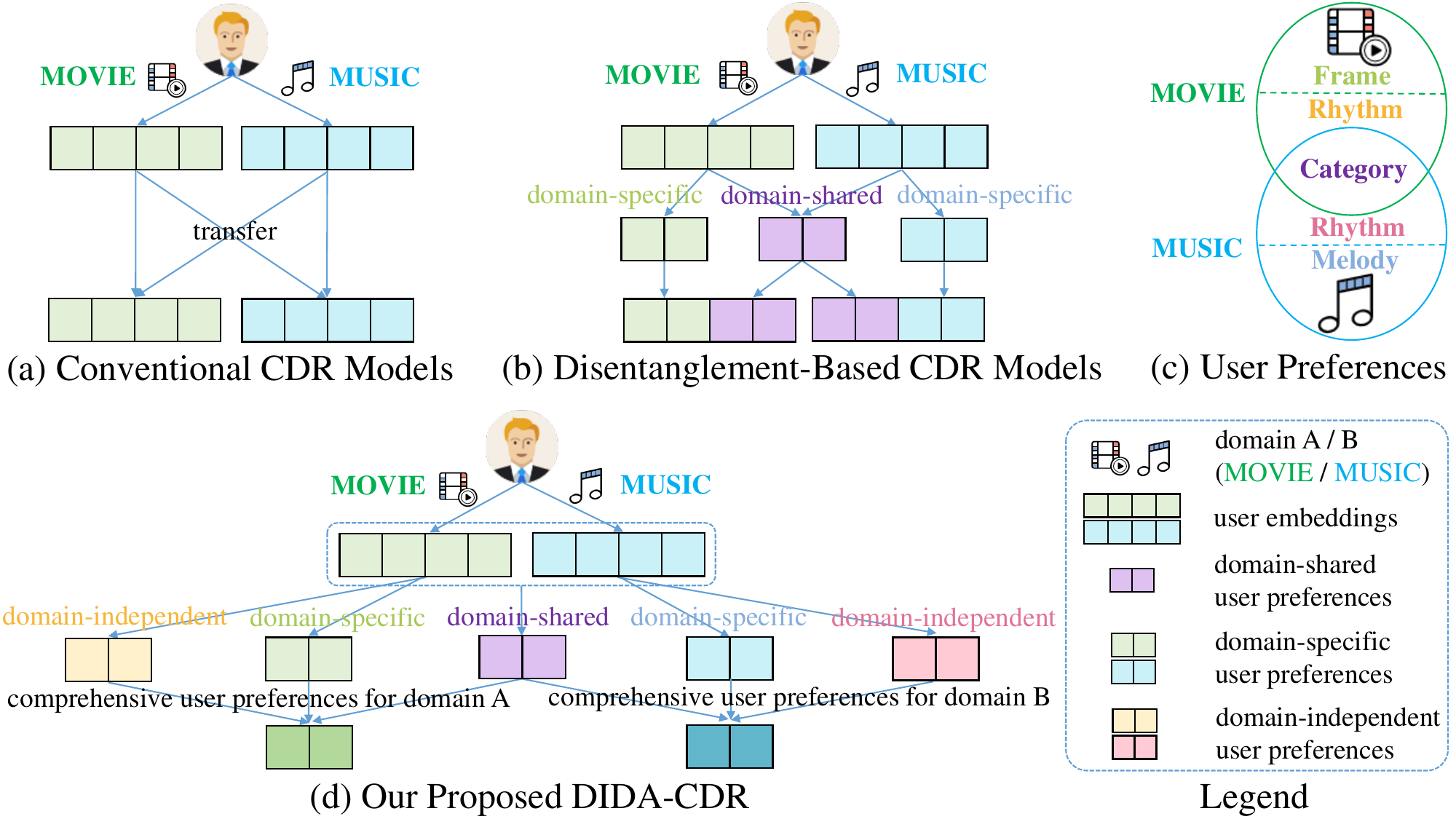} 
\caption{(a)-(b): Existing dual-target CDR models. (c): A motivating example of user preferences in movie and music domains. (d): Compared with existing dual-target CDR models, our proposed DIDA-CDR not only decouples domain-shared user preferences for cross-domain knowledge transfer, but also takes domain-independent user preferences into consideration.}
\end{figure*}

The existing dual-target CDR methods can be divided into two groups, i.e., (1) conventional methods, (2) disentanglement-based methods. \emph{Conventional methods} (see Fig. 1(a)) mainly utilize various transfer layers \cite{hu2018conet,li2020ddtcdr, zhao2019cross,liu2020cross} to integrate the representations learned by two base encoders in their respective domains. In contrast, \emph{disentanglement-based methods} (see Fig. 1(b)) tend to use the variational autoencoder (VAE) \cite{cao2022disencdr} or other disentangling techniques \cite{guo2023disentangled,zhang2022multi,cao2022cross} to decouple the domain-shared and domain-specific information, and only transfer the domain-shared information to each domain, which enhances the recommendation accuracy on both domains simultaneously. 

However, the existing dual-target CDR models have limitations in terms of effectively capturing comprehensive user preferences for the following reasons. Firstly, even though some of them use data augmentation to mitigate data imbalance between richer and sparser domains, few of them can balance the relevance (being able to represent user preferences in both domains) and diversity (having enough variations in user representations) of the user-item interaction augmentation. As a result, such augmented user representations can hardly provide strong support for subsequently capturing accurate and comprehensive user preferences. Secondly, none of the existing methods decouple all three essential components needed to capture comprehensive user preferences. To be specific, the existing methods only decouple two essential components, i.e.,  domain-shared and domain-specific information, and ignore the existence of domain-independent information. Since domain-independent information has different meanings from the other two types of information, it cannot be ignored when capturing comprehensive user preferences. It is worth mentioning here that some existing works also decouple so called ‘domain-independent information’, but its meaning is different from that in our work (cf. the detailed description in Section 2.1).

Below we introduce domain-shared, domain-specific and domain-independent information respectively with examples, and further differentiate them.

\begin{itemize}[leftmargin=*]
\item[(i)] \textbf{Domain-shared Information:} As shown in Fig. 1(c), there exists some \emph{domain-shared} information in both movie domain and music domain, such as ‘Category’. For example, people who like watching \emph{suspense movies} (i.e., a category in movie domain) tend to like listening to \emph{suspense music} (i.e., a category in music domain), and vice versa. Since the domain-shared information can provide the valuable information for cross-domain recommendations, it needs to be first decoupled and then transferred to both domains.
\item[(ii)] \textbf{Domain-specific Information:} In contrast, there also exists some \emph{domain-specific} information in each domain. For example, the user preference for pictures in a movie (i.e., ‘Frame’ in movie domain) is not applicable in music domain because ‘Frame’ is unique to movie domain. Thus, such domain-specific information should be decoupled too, which helps improve the recommendation performance on its own domain, but it should not be transferred to another domain in order to avoid the negative transfer \cite{zhu2021unified}. 
\item[(iii)] \textbf{Domain-independent Information:} In addition, some \emph{domain-independent} information also exists in each domain, but should not be transferred to other domains. For instance, ‘Rhythm’ exists in each of movie, music and book domains. However, in movie domain, it means the use of sound effects, the speed of camera cuts, and the changes in the pace of movie scenes, etc \cite{shirahama2004video}. In music domain, it means  , etc \cite{hauser2003evolution}. In book domain, it means the fluidity of the writing and the ups and downs of the storyline, etc \cite{gibbs2015writing}. In other words, although ‘Rhythm’ is seemingly common in all three domains, it has different meanings in different domains. Hence, such information is domain-independent and should be extracted from its own domain for capturing comprehensive user preferences, but should not be transferred to other domains.
\item[(iv)] \textbf{Difference:} Different to the domain-independent information, the domain-shared information (e.g., ‘Category’) extracted from two domains expresses the same meaning in each domain. In addition, the domain-specific information, e.g., ‘Frame’ in the movie domain, exists only in its own domain. By contrast, although the domain-independent information, e.g., ‘Rhythm’, exists in each domain, it has different meanings in different domains.

\item[(v)] \textbf{Summary:} Therefore, it is vital to recognize the existence of domain-independent information and to clearly differentiate it from domain-shared and domain-specific information. More importantly, decoupling domain-independent information is crucial for capturing more comprehensive user preferences, otherwise it will cause suboptimal recommendation results. However, existing dual-target CDR methods neglect the above insights. Hence, novel dual-target CDR solutions are needed to incorporate the above insights for capturing more comprehensive user preferences.
\end{itemize}

Following the above discussions, to target superior dual-target CDR, there are two major challenges as follows.

\noindent \textbf{CH1:} \emph{How to generate relevant and diverse augmented user representations to augment sparser domain and explore potential user preferences?} The existing dual-target CDR methods \cite{guo2023disentangled,xiao2023catcl} mainly obtain augmentation views by perturbing the original data. Although these methods can increase the diversity of augmentation by increasing the types of perturbations, they cannot generate augmented user representations that represent user preferences in both domains because they perform data augmentation on the data from each domain separately without considering the correlation between data from both domains, and thus can hardly capture the potential common user preferences in both domains.

\noindent \textbf{CH2:} \emph{How to effectively decouple domain-independent information from domain-specific information, in addition to domain-shared information, to capture comprehensive user preferences on each domain, thereby improving the recommendation performance on both domains?} The existing dual-target CDR methods either ignore decoupling the domain-specific and domain-shared information \cite{hu2018conet} (\emph{Group 1}), or directly transfer the domain-shared information to only fuse with the domain-specific information in each domain, overlooking the domain-independent information \cite{cao2022disencdr} (\emph{Group 2}). The methods in \emph{Group 1} (also see Fig. 1(a)) disregard the discrimination between domain-specific and domain-shared information, which may lead to the negative transfer. The methods in \emph{Group 2} (also see Fig. 1(b)) do not differentiate the domain-independent information from domain-specific information and decouple these two types of information, which results in suboptimal recommendation results.

\noindent \textbf{Our Approach and Contributions:} To address the above challenges, we propose a novel disentanglement-based framework with interpolative data augmentation for dual-target CDR. To the best of our knowledge, this is the first work in the literature that explicitly takes domain-independent information into consideration in addition to domain-shared and domain-specific information, and decouples it to capture more comprehensive user preferences for cross-domain recommendation. The characteristics and contributions of our proposed model are summarized as follows:
\begin{itemize}
    \item We first propose a \textbf{D}isentanglement-based framework with \textbf{I}nterpolative \textbf{D}ata \textbf{A}ugmentation for dual-target \textbf{C}ross-\textbf{D}omain \textbf{R}ecommendation, called DIDA-CDR, which can augment the sparser domain, disentangle three essential components of user preferences and transfer the domain-shared user preferences of common users across domains, thus enhancing the recommendation accuracy on both domains simultaneously;
    \item  To address \textbf{CH1}, we propose an interpolative data augmentation approach to generating both relevant and diverse augmented user representations, which augments the sparser domain and explores the potential common user preferences and therefore improves the recommendation performance on both domains;
    \item  To address \textbf{CH2}, we propose a disentanglement module to effectively decouple the domain-independent and domain-specific user preferences. The disentanglement module also extracts the domain-shared user preferences from augmented user representations, which can be transferred to both domains to provide the valuable information. We then apply the attention mechanism to combine the above three essential components of user preferences to capture more comprehensive user preferences in each domain, which can improve the recommendation performance on each of both domains;
    \item Extensive experiments conducted on five real-world datasets show that our proposed DIDA-CDR outperforms the best-performing state-of-the-art baselines with an average improvement of 8.54\% and 11.10\% with respect to HR@10 and NDCG@10, respectively.
\end{itemize}

\section{Related Work}
In this section, we first review the relevant literature on two major categories of CDR in Section 2.1. Next, since we utilize disentangled representation learning and interpolative data augmentation for our model, we also review the relevant literature on them in Section 2.2 and Section 2.3, respectively \cite{zhu2019dtcdr}.

\subsection{Single-Target and Dual-Target CDR}
\noindent \textbf{Single-Target CDR.} The existing single-target CDR methods can be divided into two categories, i.e., (1) content-based transfer methods and (2) feature-based transfer methods \cite{ijcai2021p639}. Content-based transfer methods \cite{kanagawa2019cross} mainly use the various content information, such as user/item attributes, tags, reviews, etc., to link domains and share their information across domains. By contrast, feature-based transfer methods \cite{hu2019transfer,fu2019deeply} aim to first get embeddings or rating patterns \cite{ijcai2019p587} using various learning techniques, and then transfer them across domains.

\noindent \textbf{Dual-Target CDR.} In contrast to single-target CDR, dual-target CDR aims to achieve better recommendation performance on both domains simultaneously, which can be extended to multiple domains, leading to multi-target CDR \cite{cui2020herograph,zhu2020graphical,guo2023disentangled}. The existing dual-target CDR approaches can be divided into two categories, i.e., (1) conventional approaches, (2) disentanglement-based approaches. For \emph{conventional dual-target CDR}, CoNet \cite{hu2018conet} utilizes cross-connection networks to achieve knowledge transfer between two domains. DDTCDR \cite{li2020ddtcdr} employs a latent orthogonal mapping function to transfer user embeddings across domains. PPGN \cite{zhao2019cross} constructs a cross-domain interaction graph to learn and transfer representations. BiTGCF \cite{liu2020cross} first leverages two base graph encoders to learn user/item embeddings, and then performs the feature propagation and transfer to fuse user embeddings. However, most of them lead to the negative transfer, because they neglect to decouple the domain-shared and domain-specific information.

\begin{table*}
\newcommand{\tabincell}[2]{\begin{tabular}{@{}#1@{}}#2\end{tabular}}
\centering
\caption{Important notations.}
\label{tab:1}
\begin{tabular}{|c|c|c|c|}
\hline
\textbf{Symbol} & \multicolumn{1}{c|}{\textbf{Definition}}                      & \textbf{Symbol} & \multicolumn{1}{c|}{\textbf{Definition}}                      \\ \hline
$k$           & the dimension of embedding matrix                                                 & $P$             & the predicted domain probability          \\ \hline
$m$         & the number of users                                             &  $O$             & the ground truth domain label                                    \\ \hline
$n$             & the number of items                             &      $\mathbf{E}_u$             & the graph embedding matrix of users                       \\ \hline
$R$            & the rating matrix                           & $\mathbf{E}_v$           & the graph embedding matrix of items               \\ \hline
$\mathcal{U}$           & the set of users                             & ${\mathbf{Z}_{sha}}$             & the domain-shared user preferences                      \\ \hline
$\mathcal{V}$           & the set of items                    & ${\mathbf{Z}_{spe}}$              & the domain-specific user preferences                    \\ \hline
$y$             & the user-item interaction & ${\mathbf{Z}_{ind}}$             & the domain-independent user preferences \\ \hline
${*^{aug}}$              & the notation for data augmentation                             & $\mathbf{E}_u^*$            & the comprehensive user preferences                           \\ \hline
${*^A}$, ${*^B}$           & \tabincell{c}{the notations for domains $A$ and $B$, e.g., ${n^A}$\\represents the number of items on domain $A$}                                 & $\mathbf{E}_u^{aug}$              & \tabincell{c}{the augmented user representations\\of common users}                                        \\ \hline
$\hat *$            & \tabincell{c}{the predicted notations, e.g., ${\hat y_{ij}}$ represents\\the predicted interaction of user ${u}_i$ on item ${v}_j$}                                                & $G$             & \tabincell{c}{the heterogeneous graph, where $Q$\\is the set of user-item relationships}                                \\ \hline
\end{tabular}%
\end{table*}

In addition, we further classify the existing \emph{disentanglement-based dual-target CDR approaches} into two classes, i.e., (1) VAE-based approaches and (2) other approaches, according to disentangling techniques. VAE-based approaches \cite{wang2022disentangled} (\emph{Class 1}) employ the explicit reconstruction loss included in the Evidence Lower Bound (ELBO) and extra regularizers as the disentanglement loss to learn desirable disentangled representations. Other approaches (\emph{Class 2}) often utilize Graph Convolutional Networks (GCNs) \cite{wang2022inter}, adversarial learning \cite{choi2022based,zhao2022multi}, self-supervised learning \cite{li2023one,guo2023disentangled,zhang2023disentangled} and fixed or flexible combination strategies \cite{zhu2019dtcdr,zhu2020graphical, xiao2023catcl} to disentangle the latent knowledge, such as domain-shared and domain-specific information. The domain-shared information is also termed as domain-invariant information in \cite{liu2019jscn,su2022cross}, domain-common information in \cite{choi2022based} and domain-independent information in \cite{liu2020exploiting,sahu2020knowledge} (note that the domain-independent information in other works does not have the same meaning as in our work). However, none of them can effectively capture comprehensive user preferences, because they entangle the domain-independent information with domain-specific information.

\subsection{Disentangled Representation Learning}
Disentangled representation learning is originally introduced in the field of computer vision \cite{chen2018isolating,gonzalez2018image} and is mainly used to learn visual features such as shape, color and location features of objects \cite{zhang2020content}. In addition to computer vision, in recent years, disentangled representation learning has also been used in recommender systems (RSs) \cite{ma2019learning,wang2020disentangled,ma2020disentangled}. The main idea of disentangled representation learning is to focus on decomposing the latent factors behind the observed instances in the low-dimension vector space \cite{bengio2013representation,wang2022disentangledsurvey}. For recommendation, MacridVAE \cite{ma2019learning} performs the macro disentanglement and micro disentanglement based on the user behavior data. DGCF \cite{wang2020disentangled} leverages a graph disentangling module to decouple user embeddings learned from user-item interaction data into fine-grained user intents. The work proposed in \cite{ma2020disentangled} reconstructs the embedding of future sequence by self-supervised learning to decouple the intentions of users.

\subsection{Interpolative Data Augmentation}
Mixup \cite{zhang2017mixup} has been recently proposed as a representative interpolative data augmentation method that enhances the prediction accuracy of networks for image classification. In addition to image classification, interpolative data augmentation has been used in recent RSs \cite{huang2021mixgcf,chen2022graph}. The general idea of interpolative data augmentation is to linearly interpolate inputs and model targets of random samples. For recommendation, existing methods with interpolative data augmentation \cite{bian2022relevant} tend to generate natural augmented user representations to model sequential user behaviors \cite{hou2022towards}.

\section{The Proposed Model}
In this section, first, we formulate the target problem of our proposed model. Next, we present a novel \textbf{D}isentanglement-based framework with \textbf{I}nterpolative \textbf{D}ata \textbf{A}ugmentation for dual-target \textbf{C}ross-\textbf{D}omain \textbf{R}ecommendation, called DIDA-CDR. Finally, we elaborate on the basic components of DIDA-CDR.

\subsection{Problem Formulation}
For the sake of better readability, we list the important notations of this paper in Table 1. This paper considers the dual-target CDR on two domains ${D^A}$ and ${D^B}$ with a shared user set, denoted by $\mathcal{U}$ (of size $m = |\mathcal{U}|$). The sets of items in ${D^A}$ and ${D^B}$ are defined as ${\mathcal{V}^A}$ (of size ${n^A} = |{\mathcal{V}^A}|$) and ${\mathcal{V}^B}$ (of size ${n^B} = |{\mathcal{V}^B}|$), respectively. Let ${\mathbf{R}^A} \in {\{ 0,1\} ^{m \times {n^A}}}$ and ${\mathbf{R}^B} \in {\{ 0,1\} ^{m \times {n^B}}}$ denote the binary user-item interaction matrices in ${D^A}$ and ${D^B}$, respectively. By aggregating the interaction data in each domain, we first construct two heterogeneous graphs ${G^A} = (\mathcal{U},{\mathcal{V}^A},{\mathcal{Q} ^A})$ and ${G^B} = (\mathcal{U},{\mathcal{V}^B},{\mathcal{Q} ^B})$ to learn user embeddings $\mathbf{E}_u^A$, $\mathbf{E}_u^B$ and item embeddings $\mathbf{E}_v^A$, $\mathbf{E}_v^B$ in domains ${D^A}$ and ${D^B}$, respectively, where ${\mathcal{Q} ^A}$ and ${\mathcal{Q} ^B}$ are the edge sets that represent the observed user-item interactions. By linearly interpolating the user embeddings $\mathbf{E}_u^A$ and $\mathbf{E}_u^B$, we then generate augmented user representations $\mathbf{E}_u^{aug}$ to augment the sparser domain ${D^B}$. Given two coarse user embeddings $\mathbf{E}_u^A$, $\mathbf{E}_u^B$ and the augmented user representations $\mathbf{E}_u^{aug}$, our goal is to disentangle domain-shared, domain-specific and domain-independent user preferences, i.e., ${\mathbf{Z}_{sha}}$, ${\mathbf{Z}_{spe}}$, and ${\mathbf{Z}_{ind}}$, and then transfer domain-shared user preferences ${\mathbf{Z}_{sha}}$ to each domain to capture comprehensive user preferences $\mathbf{E}_u^*$, thereby improving the recommendation performance on both domains.

\begin{figure*}[ht]
\centering 
\includegraphics[scale=0.51]{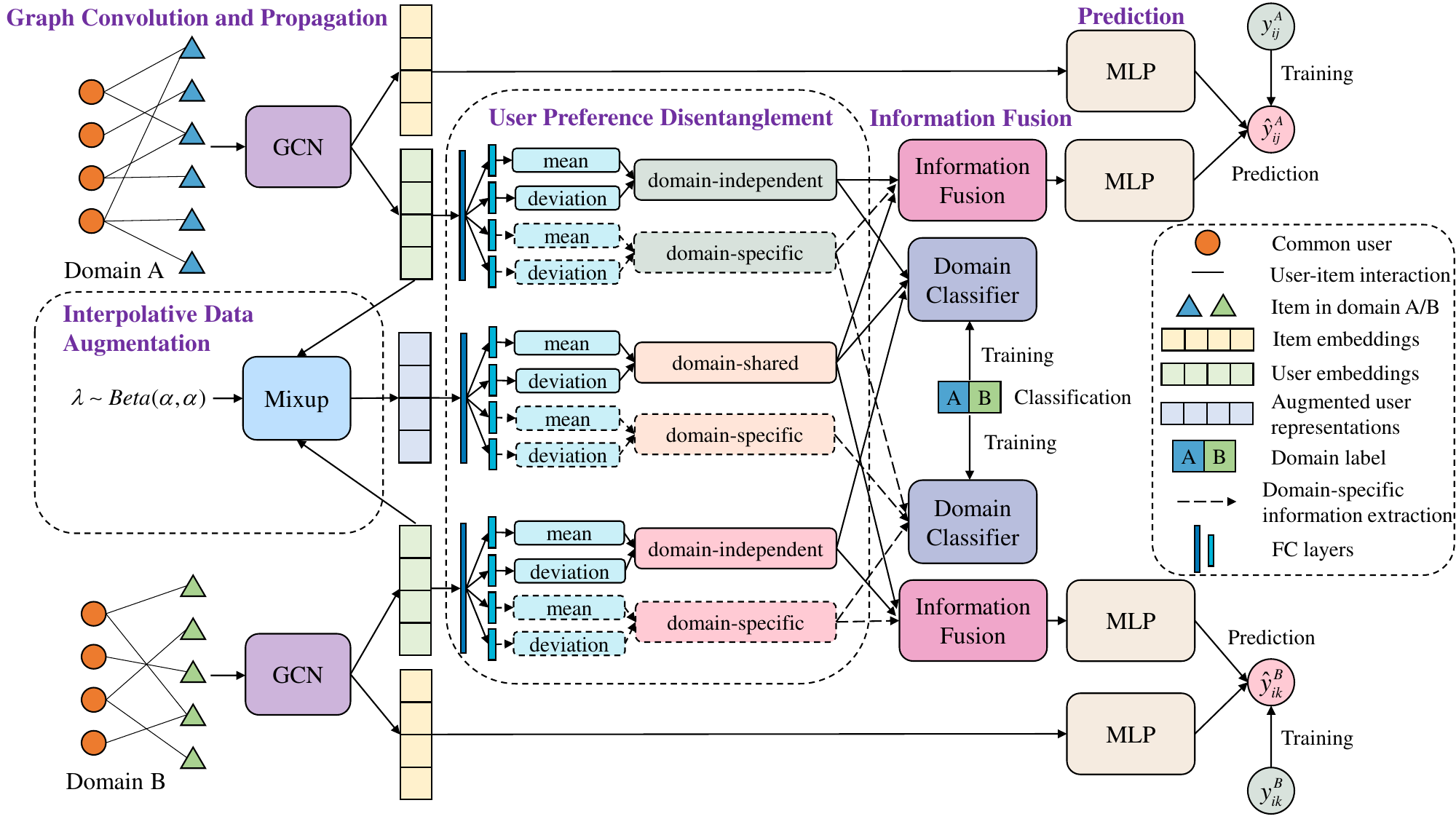}
\caption{The details of each module of our proposed DIDA-CDR.}
\end{figure*}

\subsection{Overview of DIDA-CDR}
To enhance the recommendation accuracy on both domains, we propose a novel \underline{d}isentanglement-based framework with \underline{i}nterpolative \underline{d}ata \underline{a}ugmentation for dual-target \underline{c}ross-\underline{d}omain \underline{r}ecommendation, called DIDA-CDR. This framework contains five major components, i.e., (1) \emph{Graph Convolution and Propagation Module}, (2) \emph{Interpolative Data Augmentation Module}, (3) \emph{User Preference Disentanglement Module}, (4) \emph{Information Fusion Module}, and (5) \emph{Prediction Module}. The details of each module of our proposed DIDA-CDR are illustrated in Fig. 2. They are briefly introduced below and described in detail in the following subsections.

\noindent \textbf{(1) Graph Convolution and Propagation.} First, we construct two heterogeneous graphs to extract the high-order user-item interaction relationships using the interaction data in domain $A$ and domain $B$, respectively. Based on the above graphs, we apply the graph convolution and propagation layer in the GCN \cite{KipfW17} to generate user and item embeddings.

\noindent \textbf{(2) Interpolative Data Augmentation.} Next, we propose an interpolative data augmentation approach to augmenting user embeddings at the representation level. The interpolative data augmentation approach linearly interpolates user embeddings in domain $A$ and domain $B$ to generate both relevant and diverse augmented user representations.

\noindent \textbf{(3) User Preference Disentanglement.} Thereafter, we propose a disentanglement module guided by a domain classifier to decouple more accurate domain-specific and domain-independent user preferences from user embeddings generated by GCNs. This module also disentangles the domain-shared user preferences from the augmented user representations, which are then transferred to both domains to provide the valuable information.

\noindent \textbf{(4) Information Fusion.} Next, we use three approaches, i.e., concatenation, element-wise sum, and attention mechanism to incorporate essential components of user preferences, i.e., domain-shared, domain-specific and domain-independent information, which are decoupled by the disentanglement module, to capture comprehensive user preferences.

\noindent \textbf{(5) Prediction.} Finally, we apply the multi-layer perception (MLP) to model the user-item interaction relationships, endowing the non-linearity to our proposed model. Based on the MLP, the predicted user-item interaction matrix can be obtained. The prediction loss between it and the observed user-item interaction matrix, together with two domain classification losses, constitute the final loss for training.

Overall, our model can be easily extended to a multi-target CDR model. Specifically, we can disentangle domain-shared user preferences and transfer them to all domains, decouple the domain-specific and domain-independent user preferences from each domain, and then capture the comprehensive user preferences to improve the recommendation performance on multiple domains simultaneously.

\subsection{Graph Convolution and Propagation}
GCNs excel at capturing the relationships between nodes and learning the representation of graph data, and are well suited for modeling user-item relationships in RSs, because the user-item interaction data can be easily transformed into the graph structure. To distill the high-order user-item interaction relationships, we construct two heterogeneous graphs ${G^A}$ and ${G^B}$ for domains $A$ and $B$, respectively, where nodes refer to entities (i.e., users and items) and edges refer to interactions. In this paper, we apply the graph convolution and propagation layer in the GCN to encode the user and item embeddings according to the user-item interaction matrix ${\mathbf{R}^A}$ (or ${\mathbf{R}^B}$). The node embeddings $\mathbf{E}_0^A$ (or $\mathbf{E}_0^B$) are randomly initialized. Given a graph ${G^A}$, the propagation rule is represented as:
\begin{equation}
\mathbf{E}_l^A = f({\mathbf{\tilde D}^{ - {\textstyle{1 \over 2}}}}{\mathbf{\tilde R}^A}{\mathbf{\tilde D}^{ - {\textstyle{1 \over 2}}}}\mathbf{E}_{l - 1}^A{\mathbf{W}_l} + {\mathbf{b}_l}),
\end{equation}
where ${\mathbf{\tilde R}^A} = {\mathbf{R}^A} + \mathbf{I}$ is the user-item interaction matrix of graph ${G^A}$ after adding a self-loop identity matrix $\mathbf{I}$. $\mathbf{\tilde D}$ is a degree matrix for normalization. ${\mathbf{W}_l}$ and ${\mathbf{b}_l}$ are the trainable weight matrix and bias vector in the ${l^{th}}$ layer respectively, and $\mathbf{E}_l^A$ is the hidden embedding matrix of graph ${G^A}$ in the ${l^{th}}$ layer \cite{meng2021graph}. $f( \cdot )$ denotes the ReLU activation function.

After $l$ times propagation, we can obtain the global hidden representations ${\mathbf{E}^A}$ by concatenating multiple embedding matrices from $\mathbf{E}_0^A$ to $\mathbf{E}_l^A$, which can be rearranged into the user embeddings $\mathbf{E}_u^A$ and item embeddings $\mathbf{E}_v^A$ in domain $A$ \cite{zhao2019cross}. Similarly, we can obtain the user embeddings $\mathbf{E}_u^B$ and item embeddings $\mathbf{E}_v^B$ in domain $B$.

\subsection{Interpolative Data Augmentation}
Although the user embeddings in both domains are obtained by GCNs, the user embeddings learned from the sparser domain are not as accurate as those learned from the richer domain due to the data imbalance between the two domains. To augment the sparser domain and explore the potential user preferences, inspired by the mixup technique \cite{zhang2017mixup}, we design an interpolative data augmentation approach, which generates both relevant and diverse augmented user representations. However, the conventional mixup technique cannot be directly utilized for our task, because the interaction data cannot be directly mixed at the pixel level like images. Therefore, we propose to linearly interpolate the embeddings of common users in both domains. The augmentation of common users ensures that augmented user representations maintain the relevance of user preferences in both domains, i.e., relevant augmented user representations, while corresponding to adding more interaction data for users in the sparser domain. In addition, introducing randomness in linear interpolation instead of using a fixed mixing coefficient can generate diverse augmented user representations, which can provide richer information for subsequent disentanglement. The formula for interpolative data augmentation can be expressed as follows:
\begin{equation}
\mathbf{E}_u^{aug} = \lambda \mathbf{E}_u^A + (1 - \lambda )\mathbf{E}_u^B,
\end{equation}
where $\mathbf{E}_u^{aug}$ denotes the augmented user representations\footnote{The two terms, i.e., embedding and representation, are exchangeable in this paper.}. Since we aim to generate more diverse augmented user representations through data augmentation, instead of using attention-based methods in this module, we propose the interpolative data augmentation approach. $\lambda  \in [0,1]$ is the mixing coefficient sampled from $Beta(\alpha ,\alpha )$, $\alpha  \in (0,\infty )$. The advantages of adopting a mixing coefficient sampled from $Beta(\alpha ,\alpha )$ are as follows. First, since the user embeddings of the common user in each of the two domains should be equivalent, the mixing coefficient should be sampled within the interval $[0,1]$ and be symmetric around 0.5. Beta distribution $Beta(\alpha ,\alpha )$ satisfies this characteristic. Second, it has been proven to effectively improve the generalization ability of the model \cite{wang2021mixup}. Meanwhile, $\lambda $ also introduces randomness into our model, thus weakening the negative transfer that may result from performing a linear interpolation operation with fixed weights (cf. the experimental results in Section 4.3 and the discussion of parameter $\alpha$ in Section 4.6.2).

\subsection{User Preference Disentanglement}
To capture domain comprehensive preferences of users, we utilize disentangled representation learning to extract the essential components of user preferences from previously obtained user embeddings. Inspired by the method introduced in \cite{fu2021meta}, we design a disentanglement module to decouple the domain-specific and domain-independent user preferences. Specifically, our disentanglement module adopts a similar architecture to the encoder of VAE, but it is quite different from VAE. In particular, VAE only encodes one latent feature, while our disentanglement module aims to better learn to decouple the domain-specific and domain-independent user preferences. Although both domain-specific and domain-independent user preferences cannot be transferred to other domains, they have different meanings, thus their importance in capturing comprehensive user preferences is different. If they are not distinguished, they are equally important in capturing comprehensive user preferences, which is inappropriate. By contrast, if they are decoupled, the subsequent information fusion module can use attention mechanism to learn their weights respectively, thereby capturing more accurate and comprehensive user preferences. This module can also disentangle the domain-shared user preferences from the augmented user representations. In this case, our model can not only extract the domain-shared preferences of common users in both domains, but also explore the domain-specific personalized user preferences and domain-independent user preferences, which enhances the comprehensiveness of capturing user preferences, and thus improves the recommendation performance on both domains. To this end, we first feed user embeddings in both domains (see methods introduced in Section 3.3) and augmented user representations (see methods introduced in Section 3.4) into this module and perform the following processing respectively.

Taking domain $A$ as an example, the user embeddings $\mathbf{E}_u^A$ are first entered into the disentanglement module, which consists of several fully connected (FC) layers. More specifically, the first FC layer (see navy blue FC layer in Fig. 2), followed by the ReLU activation function, is utilized to extract general representations. In addition, the subsequent FC layers (also see lake blue FC layers in Fig. 2) outputs two sets of latent vectors, each representing different mean and deviation information of the input user embeddings. Next, following the method introduced in \cite{fu2021meta}, a reparametrization trick is adopted to generate domain-independent $\mathbf{Z}_{ind}^A$ and domain-specific user preferences $\mathbf{Z}_{spe}^A$. Similarly, we can obtain the domain-independent $\mathbf{Z}_{ind}^B$ and domain-specific user preferences $\mathbf{Z}_{spe}^B$ in domain $B$, as well as the domain-shared $\mathbf{Z}_{sha}^{aug}$ and domain-specific user preferences $\mathbf{Z}_{spe}^{aug}$ decoupled from augmented user representations. 

To ensure that the above essential components of user preferences can be accurately disentangled, we introduce a domain classifier ${H_{cls}}( \cdot )$, which includes a single FC layer to predict the domain probability of user preferences, to guide the disentanglement process. In order to supervise the optimization process of disentanglement module, we further set two domain classification tasks to train our DIDA-CDR. First of all, the disentanglement module is guided to decouple the domain-specific information with stronger domain identification ability, i.e., more accurate domain-specific user preferences, by minimizing the domain classification loss ${\mathcal L_{cl{s_1}}}$. In other words, if the decoupled user preferences can be easily recognized by a domain classifier as belonging to a particular domain, then such user preferences are considered as the domain-specific information. By contrast, the domain-independent and domain-shared information are leveraged to confuse the domain classifier, i.e., to make the domain classifier unable to identify the domain to which they belong, to ensure that they can be distinguished from the domain-specific information. In other words, if the decoupled user preferences are no better than random guesses in identifying the domain to which they belong when fed into a domain classifier, then such user preferences are mutually exclusive with the domain-specific information and should be classified as the domain-independent information. When the input to the disentanglement module is the augmented user representations containing the user preferences from both domains, the above user preferences refer to the domain-shared information. Specifically, the domain classification loss ${\mathcal L_{cl{s_1}}}$ can be defined as follows:

\begin{equation}
\begin{split}
{\mathcal L_{cl{s_1}}} &= \frac{1}{3}\sum {[{\ell _1}({H_{cls}}(\mathbf{Z}_{spe}^A),O_{spe}^A)} \\
 &+ {\ell _1}({H_{cls}}(\mathbf{Z}_{spe}^B),O_{spe}^B)\\
 &+ \lambda  \cdot {\ell _1}({H_{cls}}(\mathbf{Z}_{spe}^{aug}),O_{spe}^{augA})\\
 &+ (1 - \lambda ) \cdot {\ell _1}({H_{cls}}(\mathbf{Z}_{spe}^{aug}),O_{spe}^{augB})],
\end{split}
\end{equation}
where ${\ell _1}(P,O)$ denotes the cross-entropy loss function. $P$ is the predicted domain probability of input user preferences and $O$ is the ground truth (GT) domain label. We define the corresponding GT for domain-specific user preferences $\mathbf{Z}_{spe}^A$, $\mathbf{Z}_{spe}^B$ and $\mathbf{Z}_{spe}^{aug}$ as $O_{spe}^A$, $O_{spe}^B$ and $O_{spe}^{aug}$, respectively. All the elements in $O_{spe}^A$ and $O_{spe}^{augA}$ are set to 1, while those in $O_{spe}^B$ and $O_{spe}^{augB}$ are set to 0. Here, $\lambda$ represents the confidence score that an augmented user representation belongs to its initial GT, since augmented user representations are generated by incorporating embeddings of common users in both domains with a mixing coefficient $\lambda $ \cite{fu2021meta}.

Similarly, we define the corresponding GT for domain-independent user preferences $\mathbf{Z}_{ind}^A$, $\mathbf{Z}_{ind}^B$ and domain-shared user preferences $\mathbf{Z}_{sha}^{aug}$ as $O_{ind}^A$, $O_{ind}^B$ and $O_{sha}^{aug}$, respectively. We set all the items in $O_{ind}^A$, $O_{ind}^B$ and $O_{sha}^{aug}$ as $[0.5,0.5]$ \cite{fu2022generalized}. This ensures that the learned domain-shared and domain-independent user preferences cannot be used to identify the domain to which they belong, and thus be distinguished from the domain-specific user preferences. Specifically, the loss function is represented as follows:
\begin{equation}
\begin{split}
{\mathcal L_{cl{s_2}}} &= \frac{1}{3}\sum {[{\ell _2}({H_{cls}}(\mathbf{Z}_{ind}^A),O_{ind}^A)} \\
 &+ {\ell _2}({H_{cls}}(\mathbf{Z}_{ind}^B),O_{ind}^B)\\
 &+ {\ell _2}({H_{cls}}(\mathbf{Z}_{sha}^{aug}),O_{sha}^{aug})],
\end{split}
\end{equation}
where ${\ell _2}(P,O)$ is the Kullback-Leibler divergence loss function.

\begin{table}[t]
\caption{Comparison of information fusion approaches \cite{sun2019multi}.}
\resizebox{\linewidth}{!}{
\begin{tabular}{c|c}
\toprule
\hline
                           & Formula \\ \hline
Concatenation              & $\mathbf{E}_u^* = [{\mathbf{Z}_{spe}},{\mathbf{Z}_{ind}},{\mathbf{Z}_{sha}}]$       \\ \hline
Element-wise sum           & $\mathbf{E}_u^* = {\mathbf{Z}_{spe}} + {\mathbf{Z}_{ind}} + {\mathbf{Z}_{sha}}$       \\ \hline
\multirow{2}{*}{Attention} & ${\mathbf{C}_u} = Softmax ({\mathbf{W}_s} \cdot \sigma ({\mathbf{W}_{spe}} \cdot {\mathbf{Z}_{spe}} + {\mathbf{W}_{ind}} \cdot {\mathbf{Z}_{ind}} + {\mathbf{W}_{sha}} \cdot {\mathbf{Z}_{sha}}))$      \\
                           & $\mathbf{E}_u^* = [{\mathbf{Z}_{spe}},{\mathbf{Z}_{ind}},{\mathbf{Z}_{sha}}] \cdot {\mathbf{C}_u}$       \\ \hline
\bottomrule
\end{tabular}
}
\end{table}

\subsection{Information Fusion}
The domain-shared, domain-specific and domain-independent information are three essential components of user preferences, which need to be integrated in a reasonable and efficient way to capture comprehensive user preferences. To this end, in this paper, we leverage three approaches, i.e., concatenation, element-wise sum, and attention mechanism, to aggregate individual representations into comprehensive user preferences \cite{sun2019multi}. The specific operations of these information fusion approaches are expressed in Table 2. (cf. the experimental comparison of them in Section 4.5).

\subsection{Model Prediction and Training}

After the information fusion, we obtain the comprehensive user preferences $\mathbf{E}_u^*$ and we also have the corresponding item embeddings $\mathbf{E}_v$ generated by GCNs. To give our model the non-linearity, we adopt a neural network, i.e., MLP, to represent the user-item interactions. Taking the domain $A$ as an example, the input user embeddings and item embeddings in domain $A$ for the MLP are defined as $\mathbf{S}_{in}^A = \mathbf{E}_u^{A*}$ and $\mathbf{T}_{in}^A = \mathbf{E}_v^A$, respectively. Moreover, the output embeddings of user ${u_i}$ and item ${v_j}$ of MLP is expressed as:
\begin{equation}
\mathbf{S}_i^A = \mathbf{S}_{ou{t_i}}^A = \delta( \ldots \delta(\delta(\mathbf{S}_{i{n_i}}^A \cdot W_{{\mathbf{S}_1}}^A) \cdot W_{{\mathbf{S}_2}}^A)),
\end{equation}
\begin{equation}
\mathbf{T}_i^A = \mathbf{T}_{ou{t_i}}^A = \delta( \ldots \delta(\delta(\mathbf{T}_{i{n_i}}^A \cdot W_{{\mathbf{T}_1}}^A) \cdot W_{{\mathbf{T}_2}}^A)),
\end{equation}
where $\delta( \cdot )$ is the LeakyReLU activation function. $\mathbf{W}_{{\mathbf{S}_1}}^A$, $\mathbf{W}_{{\mathbf{S}_2}}^A \ldots$ and $\mathbf{W}_{{\mathbf{T}_1}}^A$, $\mathbf{W}_{{\mathbf{T}_2}}^A \ldots$ denote the trainable weight matrices of MLP in various layers, respectively. 

Next, the predicted user-item interaction $\hat y_{ij}^A$ between user ${u_i}$ and item ${v_j}$ in domain $A$ can be formulated as follows:
\begin{equation}
\hat y_{ij}^A = cosine(\mathbf{S}_i^A,\mathbf{T}_j^A) = \frac{{\mathbf{S}_i^A \cdot \mathbf{T}_j^A}}{{\| \mathbf{S}_i^A \| \| \mathbf{T}_j^A \|}}.
\end{equation}
Furthermore, the prediction loss in domain $A$ is defined as follows:
\begin{equation}
{\mathcal L_{prd}^A} = \sum\limits_{y \in {\mathcal{Y}^{A+}} \cup {\mathcal{Y}^{A-}}} {\ell _1 (\hat y,y)}  + \gamma (\| {{\mathbf{S}^A}} \|_F^2 + \| {{\mathbf{T}^A}} \|_F^2),
\end{equation}
where ${\ell _1}(\hat y,y)$ is the cross-entropy loss function. $y$ denotes an observed user-item interaction, and $\hat y$ is the corresponding predicted user-item interaction. ${{\mathcal{Y}^{A+}}}$ denotes the set of observed interactions and ${{\mathcal{Y}^{A-}}}$ is a certain number of negative instances randomly sampled from the set of unseen interactions in domain $A$ to avoid over-fitting. $\| {{\mathbf{S}^A}} \|_F^2 + \| {{\mathbf{T}^A}} \|_F^2$ is a regularizer controlled by $\gamma$.

Finally, we utilize a multi-task learning mechanism consisting of a prediction task and two domain classification tasks to optimize our model in domain $A$. Specifically, the final loss function is formulated as follows:
\begin{equation}
{\mathcal L^A} = {\mathcal L_{prd}^A} +  {\mu _1} \cdot {\mathcal L_{cl{s_1}}} +  {\mu _2} \cdot {\mathcal L_{cl{s_2}}}.
\end{equation}
where ${\mu _1}$ and ${\mu _2}$ denote the weights of domain classification losses ${\mathcal L_{cl{s_1}}}$ and ${\mathcal L_{cl{s_2}}}$, respectively (cf. the discussion of parameters ${\mu _1}$ and ${\mu _2}$ in Section 4.6.3). Similarly, we perform the same optimization process for domain $B$.

\section{Experiments and Analysis}
In order to demonstrate the superiority of our proposed DIDA-CDR and explore the effectiveness of its various modules, we conduct extensive experiments on five real-world datasets to answer the following five research questions:

\begin{itemize}[leftmargin=*]
    \item \noindent \textbf{RQ1.} How does our model perform when compared to representative and state-of-the-art baseline models?
    \item \noindent \textbf{RQ2.} How do various modules (i.e., interpolative data augmentation and user preference disentanglement) affect the results of our model?
    \item \noindent \textbf{RQ3.} How do various components of user preferences (i.e., domain-shared, domain-specific and domain-independent information) contribute to the performance improvement of our model?
    \item \noindent \textbf{RQ4.} How do various information fusion approaches influence the performance of our models?
    \item \noindent \textbf{RQ5.} How does the performance of our model change with various values of hyper-parameters?
\end{itemize}

\subsection{Experimental Settings}
\subsubsection{\textbf{Experimental Datasets and Tasks.}}
In order to verify the recommendation performance of our proposed DIDA-CDR, we conduct extensive experiments on five real-world datasets, i.e., Douban subsets (Douban-Movie, Douban-Book and Douban-Music) released in GA-DTCDR \cite{zhu2020graphical} and Amazon subsets (Amazon-Elec and Amazon-Cloth) released in DisenCDR \cite{cao2022disencdr}. For these five datasets, we first convert the explicit ratings into implicit feedback, i.e., we binarize the ratings into 0 and 1 to indicate whether the user has interacted with the item or not. Following the methods introduced in \cite{liu2022exploiting,liu2020cross}, we then filter these datasets to remove users and items with less than 5 interactions. Since the two Amazon subsets are filtered out of the cold-start item entry in the test set, following DisenCDR \cite{cao2022disencdr}, we perform the same preprocessing operation on the three Douban subsets as well for a fair comparison. Finally, we divide the above five subsets into three pairs of datasets, extract the common users in each pair of datasets, and design three dual-target CDR tasks in a scenario where users completely overlap, which can be listed as follows:
\begin{itemize}
    \item \noindent \textbf{Task 1:} Douban-Movie (richer) $\leftrightarrow$ Douban-Book (sparser)
    \item \noindent \textbf{Task 2:} Douban-Movie (richer) $\leftrightarrow$ Douban-Music (sparser)
    \item \noindent \textbf{Task 3:} Amazon-Elec (richer) $\leftrightarrow$ Amazon-Cloth (sparser)
\end{itemize}
The details of these three dual-target CDR tasks and the corresponding datasets are shown in Table 3.

\begin{table}[t]
\caption{Statistics of three dual-target CDR tasks.}
\resizebox{\linewidth}{!}{
\begin{tabular}{cccccc}
\toprule
\hline
Tasks                   & Datasets     & \#Users & \#Items & \#Interactions & Density \\ \hline
\multirow{2}{*}{Task 1} & Douban-Movie & 2106    & 9555    & 907219         & 4.508\% \\
                        & Douban-Book  & 2106    & 6777    & 95974          & 0.672\% \\ \hline
\multirow{2}{*}{Task 2} & Douban-Movie & 1666    & 9555    & 781288         & 4.908\% \\
                        & Douban-Music & 1666    & 5567    & 69681          & 0.751\% \\ \hline
\multirow{2}{*}{Task 3} & Amazon-Elec  & 15761   & 51447   & 224689         & 0.027\% \\
                        & Amazon-Cloth & 15761   & 48781   & 133609         & 0.017\% \\ \hline
\bottomrule
\end{tabular}
}
\end{table}

\subsubsection{\textbf{Parameter Settings.}}
For the graph convolution and propagation module in Fig. 2, the layer structure of GCN is ‘$k$ $\rightarrow$ $k$’ and for the disentanglement module, the layer structure is ‘$2k$ $\rightarrow$ $k$’. In the prediction module, the layer structure of user branch MLP is ‘$k$ $\rightarrow$ $2k$ $\rightarrow$ $k$’, and the layer structure of item branch MLP is ‘$2k$ $\rightarrow$ $2k$ $\rightarrow$ $k$’. $k$ is the embedding dimension. We vary $k$ in the range of $\{64,128\}$, but in order to balance the trade-off of recommendation accuracy and model training time, we finally set $k$ to 64. The parameters of all these layers are initialized as the Gaussian distribution $X \sim \mathcal N(0,0.01)$. For each observed user-item interaction, following GA-DTCDR \cite{zhu2020graphical}, we randomly sample 7 unseen interactions as negative instances. For a fair comparison, we leverage the grid search to tune the choice of parameters of all models. For the baseline models, we tune them based on the best parameter settings listed in their original papers. Specifically, we choose the learning rate from $\{0.01, 0.005, 0.001, 0.0005, 0.0001\}$, and search the regularization coefficient in the range of $\{0.001, 0.0001, 0.00001\}$. In addition, we apply the Adam \cite{kingma2014adam} to optimize all the models, and the batch size is 1024. We train our model and other baseline models with 100 epochs in order to guarantee the convergence. Moreover, we investigate the number of GCN layers $l$ in $\{ 1,2,3,4\}$, $\alpha$ of $Beta(\alpha ,\alpha )$ in $\{0.1, 0.5, 1, 2, 5\}$ and the weights of domain classification losses ${\mu _1}$, ${\mu _2}$ in $\{0.1, 0.3, 0.5, 0.7, 1, 3, 5, 10\}$, and analyze their impact on the recommendation performance of our model in Section 4.6. In the experiments, we set $l = 2$, $\alpha = 1$, and ${\mu _1} = {\mu _2} = 1$ by default and we resample the mixing coefficient $\lambda $ once for training each batch of data.

\subsubsection{\textbf{Evaluation Metrics.}}
Since the leave-one-out method is ubiquitous in baseline models, such as GA-DTCDR \cite{zhu2020graphical} and  BiTGCF \cite{liu2020cross}, we also employ it to evaluate the recommendation performance of our proposed DIDA-CDR and baseline models. In other words, we utilize the last interaction record of each test user to form the test set, while all the other interaction records are used as the training set. Following the methods introduced in \cite{krichene2022sampled,cao2022disencdr}, for each test user-item interaction, we randomly sample 999 items that the test user has not interacted with as negative items, and then predict 1000 candidate scores for ranking. The leave-one-out method contains two main metrics, i.e., Hit Ratio (HR) and Normalized Discounted Cumulative Gain (NDCG) \cite{wang2019kgat}, which are widely-used ranking evaluation metrics \cite{zhu2021unified,liu2022exploiting}. In the experiments, we employ them to evaluate the performance of the top-10 ranking results. For a fair evaluation, we perform all experiments 5 times and present the average results.

\begin{table*}[]
\caption{The comparison of the baselines and our approach \cite{zhu2020graphical}.}
\resizebox{\textwidth}{!}{%
\begin{tabular}{ccc|c|c}
\toprule
\hline
&{\textbf{Model}}        &                                                                                                                                                                     & \textbf{Embedding Strategy}                 & \textbf{Transfer Strategy}           \\ \hline
\multicolumn{1}{c|}{\multirow{9}{*}{Baselines}} & \multicolumn{1}{c|}{\multirow{2}{*}{\begin{tabular}[c]{@{}c@{}}Single-Domain \\ Recommendation (SDR)\end{tabular}}}                    & \textbf{NGCF} \cite{wang2019neural}        & Non-linear MLP                              & -                                    \\ \cline{3-5} 
\multicolumn{1}{c|}{}                           & \multicolumn{1}{c|}{}                                                                                                                  & \textbf{LightGCN} \cite{he2020lightgcn}    & Non-linear MLP                              & -                                    \\ \cline{2-5} 
\multicolumn{1}{c|}{}                           & \multicolumn{1}{c|}{\multirow{2}{*}{\begin{tabular}[c]{@{}c@{}}Single-Target Cross-Domain \\ Recommendation (CDR)\end{tabular}}}       & \textbf{BPR\_EMCDR} \cite{man2017cross}  & Linear Matrix Factorization (MF)            & MLP                                  \\ \cline{3-5} 
\multicolumn{1}{c|}{}                           & \multicolumn{1}{c|}{}                                                                                                                  & \textbf{BPR\_DCDCSR} \cite{ijcai2018p516} & Linear MF                                   & Combination \& MLP                   \\ \cline{2-5} 
\multicolumn{1}{c|}{}                           & \multicolumn{1}{c|}{\multirow{2}{*}{\begin{tabular}[c]{@{}c@{}}Conventional \\ Dual-Target CDR\end{tabular}}} & \textbf{PPGN} \cite{zhao2019cross}        & Graph Embedding                             & Combination \& MLP                   \\ \cline{3-5} 
\multicolumn{1}{c|}{}                           & \multicolumn{1}{c|}{}                                                                                                                  & \textbf{BiTGCF} \cite{liu2020cross}      & Graph Embedding                             & Transfer Learning                    \\ 
\cline{2-5}
\multicolumn{1}{c|}{}                           & \multicolumn{1}{c|}{\multirow{3}{*}{\begin{tabular}[c]{@{}c@{}}Disentanglement-Based Dual-Target \\ or Multi-Target CDR\end{tabular}}}                          & \textbf{GA-DTCDR} \cite{zhu2020graphical}    & Graph Embedding                             & Combination (Element-wise Attention) \\ \cline{3-5} 
\multicolumn{1}{c|}{}                           & \multicolumn{1}{c|}{}                                                                                                                  & \textbf{DR-MTCDR} \cite{guo2023disentangled}    & Graph Embedding \& Self-supervised Learning & Domain Adaptation                    \\ \cline{3-5} 
\multicolumn{1}{c|}{}                           & \multicolumn{1}{c|}{}                                                                                                                  & \textbf{DisenCDR} \cite{cao2022disencdr}    & Graph Embedding \& VAE                      & Transfer Learning                    \\ 
\hline
\multicolumn{1}{c|}{Our Approach}               & \multicolumn{1}{c|}{\begin{tabular}[c]{@{}c@{}}Disentanglement-Based \\ Dual-Target CDR\end{tabular}}                                  & \textbf{DIDA-CDR}     & Graph Embedding \&  Disentanglement         & Transfer Learning \& Combination (Attention)              \\ \hline
\bottomrule
\end{tabular}
}
\end{table*}

\begin{table*}[]
\caption{Performance comparison (\%) of different approaches for three dual-target CDR tasks according to HR@10 and NDCG@10 \cite{zhu2020graphical}. While the results of best-performing baselines are underlined, the best results are marked in bold (* indicates $p < 0.05$, paired t-test of our proposed DIDA-CDR vs. the best-performing baselines) \cite{zhu2021learning}.}
\begin{tabular}{c|cccc|cccc|cccc}
\toprule
\hline
\multirow{3}{*}{Datasets} & \multicolumn{4}{c|}{SDR Baselines}                                 & \multicolumn{4}{c|}{Single-Target CDR Baselines}                                                                   & \multicolumn{4}{c}{\begin{tabular}[c]{@{}c@{}}Conventional Dual-Target \\ CDR Baselines\end{tabular}} \\ \cline{2-13} 
                          & \multicolumn{2}{c|}{NGCF}          & \multicolumn{2}{c|}{LightGCN} & \multicolumn{2}{c|}{\begin{tabular}[c]{@{}c@{}}BPR\_EMCDR\\ \_MLP\end{tabular}} & \multicolumn{2}{c|}{BPR\_DCDCSR} & \multicolumn{2}{c|}{PPGN}                                    & \multicolumn{2}{c}{BiTGCF}             \\ \cline{2-13} 
                          & HR    & \multicolumn{1}{c|}{NDCG}  & HR            & NDCG          & HR                           & \multicolumn{1}{c|}{NDCG}                        & HR              & NDCG           & HR                 & \multicolumn{1}{c|}{NDCG}               & HR                 & NDCG              \\ \hline
Douban-Movie              & 10.26 & \multicolumn{1}{c|}{5.37}  & 10.53         & 5.49          & -                            & \multicolumn{1}{c|}{-}                           & -               & -              & 12.03              & \multicolumn{1}{c|}{6.42}               & 12.11              & 6.46              \\
Douban-Book               & 7.31  & \multicolumn{1}{c|}{4.08}  & 7.35          & 4.15          & 6.25                         & \multicolumn{1}{c|}{3.93}                        & 6.74            & 4.02           & 10.52              & \multicolumn{1}{c|}{4.78}               & 10.58              & 4.93              \\ \hline
Douban-Movie              & 9.16  & \multicolumn{1}{c|}{4.23}  & 9.24          & 4.25          & -                            & \multicolumn{1}{c|}{-}                           & -               & -              & 10.09              & \multicolumn{1}{c|}{4.35}               & 10.14              & 4.41              \\
Douban-Music              & 6.11  & \multicolumn{1}{c|}{3.87}  & 6.36          & 3.99          & 5.08                         & \multicolumn{1}{c|}{3.45}                        & 5.97            & 3.79           & 7.24               & \multicolumn{1}{c|}{4.03}               & 7.32               & 4.10              \\ \hline
Amazon-Elec               & 20.22 & \multicolumn{1}{c|}{11.97} & 20.03         & 10.94         & -                            & \multicolumn{1}{c|}{-}                           & -               & -              & 22.06              & \multicolumn{1}{c|}{12.44}              & 21.79              & 12.31             \\
Amazon-Cloth              & 10.95 & \multicolumn{1}{c|}{6.01}  & 11.38         & 6.10          & 9.87                         & \multicolumn{1}{c|}{5.33}                        & 10.90           & 5.86           & 13.04              & \multicolumn{1}{c|}{6.91}               & 13.16              & 6.88              \\ \hline
\bottomrule
\end{tabular}%
\vspace{-0.13in}
\end{table*}

\begin{table*}[]
\resizebox{\textwidth}{!}{%
\begin{tabular}{c|cccccc|cccccccc|cc}
\toprule
\hline
\multirow{3}{*}{\textbf{Datasets}} & \multicolumn{6}{c|}{\textbf{\begin{tabular}[c]{@{}c@{}}Disentanglement-Based Dual-Target \\ or Multi-Target CDR Baselines\end{tabular}}} & \multicolumn{8}{c|}{\textbf{Disentanglement-Based Dual-Target CDR (our)}}                                                                                                                                                                                                                                             & \multicolumn{2}{c}{\textbf{Improvement}}                                                             \\ \cline{2-17} 
                                   & \multicolumn{2}{c|}{GA-DTCDR}                    & \multicolumn{2}{c|}{DR-MTCDR}                       & \multicolumn{2}{c|}{DisenCDR}   & \multicolumn{2}{c|}{\textbf{\begin{tabular}[c]{@{}c@{}}DIDA-CDR\\ \_Fixed\end{tabular}}} & \multicolumn{2}{c|}{\textbf{\begin{tabular}[c]{@{}c@{}}DIDA-CDR\\ \_Base\end{tabular}}} & \multicolumn{2}{c|}{\textbf{\begin{tabular}[c]{@{}c@{}}DIDA-CDR\\ \_ELBO\end{tabular}}} & \multicolumn{2}{c|}{\textbf{DIDA-CDR}} & \multicolumn{2}{c}{\textbf{\begin{tabular}[c]{@{}c@{}}(DIDA-CDR vs.\\ best baselines)\end{tabular}}} \\ \cline{2-17} 
                                   & HR              & \multicolumn{1}{c|}{NDCG}      & HR             & \multicolumn{1}{c|}{NDCG}          & HR             & NDCG           & HR                               & \multicolumn{1}{c|}{NDCG}                             & HR                               & \multicolumn{1}{c|}{NDCG}                            & HR                               & \multicolumn{1}{c|}{NDCG}                            & HR                 & NDCG              & HR                                                & NDCG                                             \\ \hline
Douban-Movie                       & 12.25           & \multicolumn{1}{c|}{6.51}      & 14.74          & \multicolumn{1}{c|}{7.89}          & {\underline{15.09}}    & {\underline{8.02}}     & 16.12                            & \multicolumn{1}{c|}{8.97}                             & 12.98                            & \multicolumn{1}{c|}{6.51}                            & 15.74                            & \multicolumn{1}{c|}{8.51}                            & \textbf{16.66*}    & \textbf{9.16*}    & 10.40\%                                           & 14.21\%                                          \\
Douban-Book                        & 10.71           & \multicolumn{1}{c|}{5.06}      & {\underline{12.66}}    & \multicolumn{1}{c|}{{\underline{7.45}}}    & 12.40          & 7.27           & 13.08                            & \multicolumn{1}{c|}{7.82}                             & 10.41                            & \multicolumn{1}{c|}{5.52}                            & 12.80                            & \multicolumn{1}{c|}{7.69}                            & \textbf{13.79*}    & \textbf{8.18*}    & 8.93\%                                            & 9.80\%                                           \\ \hline
Douban-Movie                       & 10.35           & \multicolumn{1}{c|}{4.57}      & 11.27          & \multicolumn{1}{c|}{5.74}          & {\underline{12.13}}    & {\underline{5.95}}     & 12.76                            & \multicolumn{1}{c|}{6.23}                             & 10.45                            & \multicolumn{1}{c|}{4.99}                            & 12.55                            & \multicolumn{1}{c|}{6.18}                            & \textbf{13.01*}    & \textbf{6.60*}    & 7.25\%                                            & 10.92\%                                          \\
Douban-Music                       & 7.42            & \multicolumn{1}{c|}{4.19}      & 8.49           & \multicolumn{1}{c|}{4.73}          & {\underline{8.92}}     & {\underline{5.02}}     & 9.54                             & \multicolumn{1}{c|}{5.49}                             & 8.36                             & \multicolumn{1}{c|}{4.24}                            & 9.36                             & \multicolumn{1}{c|}{5.37}                            & \textbf{9.97*}     & \textbf{5.79*}    & 11.77\%                                           & 15.34\%                                          \\ \hline
Amazon-Elec                        & {\underline{23.87}}     & \multicolumn{1}{c|}{13.20}     & 22.34          & \multicolumn{1}{c|}{12.98}         & 23.77          & {\underline{13.61}}    & 24.68                            & \multicolumn{1}{c|}{13.83}                            & 22.03                            & \multicolumn{1}{c|}{12.97}                           & 24.14                            & \multicolumn{1}{c|}{13.72}                           & \textbf{25.05*}    & \textbf{14.36*}   & 4.94\%                                            & 5.51\%                                           \\
Amazon-Cloth                       & 13.94           & \multicolumn{1}{c|}{7.09}      & 14.13          & \multicolumn{1}{c|}{7.59}          & {\underline{15.46}}    & {\underline{8.42}}     & 16.01                            & \multicolumn{1}{c|}{8.94}                             & 12.97                            & \multicolumn{1}{c|}{6.93}                            & 15.76                            & \multicolumn{1}{c|}{8.63}                            & \textbf{16.69*}    & \textbf{9.33}     & 7.96\%                                            & 10.81\%                                          \\ \hline
\bottomrule
\end{tabular}%
}
\end{table*}

\subsubsection{\textbf{Comparison Methods.}}
We select a total of nine representative and state-of-the-art baseline models to compare with our proposed DIDA-CDR. These nine baseline models can be divided into four categories, i.e., (1) Single-Domain Recommendation (SDR), (2) Single-Target Cross-Domain Recommendation (CDR), (3) Conventional Dual-Target CDR, and (4) Disentanglement-Based Dual-Target or Multi-Target CDR. For a clear comparison, in Table 4, we elaborate on the embedding strategies and transfer strategies of the above nine baseline models and our proposed DIDA-CDR.

\subsection{Performance Comparison (for RQ1)}
Table 5 presents the performance comparison\footnote{Due to space limitation, we only show the results when $k=64$ in Table 5. For the results under other values of $k$ that are omitted, similarly, our model also has a significant improvement over other baseline models.} of various approaches for three dual-target CDR tasks according to HR@10 and NDCG@10. Note that since single-target CDR baselines aim to improve the recommendation performance on the sparser domain, we train them on both domains and then only present their experimental results on the sparser domain. From Table 5, we have the following observations: (1) Our proposed DIDA-CDR improves other baseline models on sparser domain by a large margin. Specifically, on sparser domain, it outperforms the best-performing baseline model with an average improvement of 9.55\% in terms of HR@10 and 11.98\% in terms of NDCG@10. This is because we adopt the interpolative data augmentation, which effectively generate both relevant and diverse augmented user representations to augment the sparser domain, and therefore significant improvements in recommendation performance can be obtained on the sparser domain; (2) Disentanglement-based dual-target CDR models improve conventional dual-target CDR models by an average of 14.98\% in terms of HR@10 and 20.94\% in terms of NDCG@10, which shows that decoupling and then transferring the domain-shared information to both domains is an efficient way for dual-target CDR; (3) Compared with other disentanglement-based dual-target CDR baselines, our model can achieve better recommendation performance over them. Specifically, our proposed DIDA-CDR achieves an average increase of 8.54\% in terms of HR@10 and 11.10\% in terms of NDCG@10, compared to the best-performing disentanglement-based dual-target CDR baseline. This is because our model particularly takes domain-independent user preferences into consideration, and the proposed disentanglement module can more effectively disentangle all three essential components of user preferences, thus capturing more comprehensive user preferences. For a more detailed analysis of each module and each component of user preferences  of the proposed DIDA-CDR, please refer to Section 4.3 and Section 4.4, respectively.

\subsection{Ablation Study (for RQ2)}
To show the contribution of each proposed component to the improvement of overall performance, we modify our proposed model to form three variants and conduct an ablation study for three dual-target CDR tasks.
\subsubsection{\textbf{Impact of Interpolative Data Augmentation}}
We construct a variant of DIDA-CDR, namely \textbf{DIDA-CDR\_Fixed}, by replacing the interpolative data augmentation module with a fixed mixing strategy. In fact, we conducted experiments to select the best-performing mixing coefficient, i.e., 0.5, from $\{0.1,0.3,0.5,0.7,0.9\}$ to implement the above variant. From Table 5, we can observe that with the interpolative data augmentation, our proposed DIDA-CDR outperforms \textbf{DIDA-CDR\_Fixed} with an average improvement of 3.7\%. This demonstrates that the interpolative data augmentation can not only augment the sparser domain by introducing randomness to increase the diversity of augmentation, but also generate representative augmented user representations by effectively mixing user embeddings for subsequent disentanglement. Meanwhile, the introduction of randomness also weakens the possible negative transfer caused by the linear interpolation operation with fixed weights, which can also be seen from the experimental results.

\subsubsection{\textbf{Impact of User Preference Disentanglement}}
Furthermore, another variant, namely \textbf{DIDA-CDR\_Base}, directly feeds the generated user embeddings to the information fusion module and does not include the disentanglement module, thus the variant only includes the prediction loss ${\mathcal L_{prd}}$, that is ${\mu _1} = {\mu _2} = 0$. From Table 5, we can observe that without the disentanglement module, the recommendation performance of \textbf{DIDA-CDR\_Base} would degrade to be comparable to that of the conventional dual-target CDR baselines, and weaker than that of disentanglement-based ones. This shows that the disentanglement module can indeed help the model perform more effective cross-domain knowledge transfer without negative transfer by decoupling domain-shared, domain-specific, domain-independent information and transfering only domain-shared information, thus improving the performance of cross-domain recommendations.

\subsubsection{\textbf{Impact of Domain Classifier}}
In addition, following DisenCDR \cite{cao2022disencdr}, we modify the disentanglement module in our model, i.e., replace our domain classification losses with the standard ELBO, to form another variant, namely \textbf{DIDA-CDR\_ELBO}. From Table 5, we can observe that our proposed DIDA-CDR outperforms \textbf{DIDA-CDR\_ELBO} with an average improvement of 5.98\%. This demonstrates that the proposed disentanglement module can indeed collaborate well with the domain classifier to decouple more accurate essential components of user preferences, especially the domain-independent information, to capture comprehensive user preferences, thus enabling the model to achieve better recommendation performance through superior disentanglement.

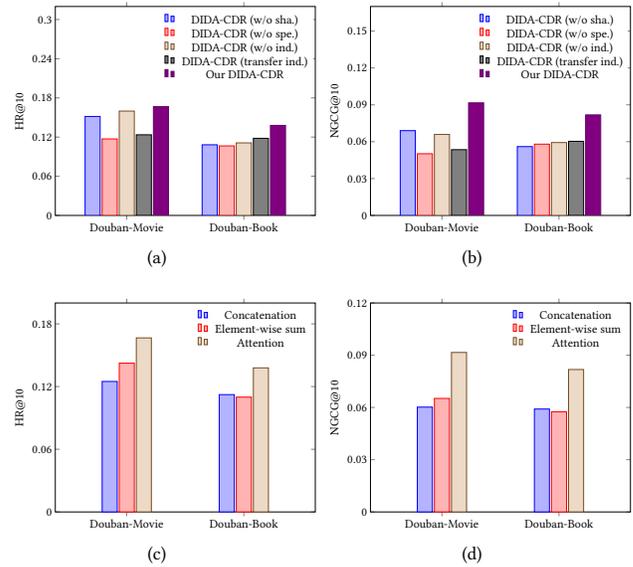
\begin{figure}[t]
\centering
\subfigure[]{
\begin{tikzpicture}[scale=0.38]
\pgfplotsset{%
    width=0.60\textwidth,
    height=0.50\textwidth
}
\begin{axis}[
    ybar,
    bar width=15pt,
    ylabel={HR@10},
    ylabel style ={font = \LARGE},
    xlabel style ={font = \LARGE},
    enlarge x limits={abs=2.5cm},
    scaled ticks=false,
    tick label style={/pgf/number format/fixed, font=\LARGE},
    ymin=0, ymax=0.32,
    symbolic x coords={Douban-Movie, Douban-Book},
    xtick=data,
    ytick={0,0.06,0.12,0.18,0.24,0.30},
    legend style={at={(0.70,0.98)}, anchor=north,legend columns=1, column sep=0.2cm, draw=none, font=\LARGE},
]
\addplot coordinates {
(Douban-Movie,0.1516) (Douban-Book,0.1082)};
\addplot coordinates {
(Douban-Movie,0.1172) (Douban-Book,0.1065)};  
\addplot coordinates {
(Douban-Movie,0.1598) (Douban-Book,0.1111)};
\addplot coordinates {
(Douban-Movie,0.1235) (Douban-Book,0.1181)};
\addplot coordinates {
(Douban-Movie,0.1666) (Douban-Book,0.1379)};
\legend{DIDA-CDR (w/o sha.), DIDA-CDR (w/o spe.), DIDA-CDR (w/o ind.), DIDA-CDR (transfer ind.), Our DIDA-CDR}
\end{axis}
\end{tikzpicture}}
\subfigure[]{
\begin{tikzpicture}[scale=0.38]
\pgfplotsset{%
    width=0.60\textwidth,
    height=0.50\textwidth
}
\begin{axis}[
    ybar,
    bar width=15pt,
    ylabel={NGCG@10},
    ylabel style ={font = \LARGE},
    xlabel style ={font = \LARGE},
    enlarge x limits={abs=2.5cm},
    scaled ticks=false,
    tick label style={/pgf/number format/fixed, font=\LARGE},
    ymin=0, ymax=0.17,
    symbolic x coords={Douban-Movie, Douban-Book},
    xtick=data,
    ytick={0,0.03,0.06,0.09,0.12,0.15},
    legend style={at={(0.70,0.98)}, anchor=north,legend columns=1, column sep=0.2cm, draw=none, font=\LARGE},
]
\addplot coordinates {
(Douban-Movie,0.069) (Douban-Book,0.056)};
\addplot coordinates {
(Douban-Movie,0.0502) (Douban-Book,0.058)};  
\addplot coordinates {
(Douban-Movie,0.0659) (Douban-Book,0.0593)};
\addplot coordinates {
(Douban-Movie,0.0535) (Douban-Book,0.0603)};
\addplot coordinates {
(Douban-Movie,0.0916) (Douban-Book,0.0818)};
\legend{DIDA-CDR (w/o sha.), DIDA-CDR (w/o spe.), DIDA-CDR (w/o ind.), DIDA-CDR (transfer ind.), Our DIDA-CDR}
\end{axis}
\end{tikzpicture}}
\subfigure[]{
\begin{tikzpicture}[scale=0.38]
\pgfplotsset{%
    width=0.60\textwidth,
    height=0.50\textwidth
}
\begin{axis}[
    ybar,
    bar width=15pt,
    ylabel={HR@10},
    ylabel style ={font = \LARGE},
    xlabel style ={font = \LARGE},
    enlarge x limits={abs=2.5cm},
    scaled ticks=false,
    tick label style={/pgf/number format/fixed, font=\LARGE},
    ymin=0, ymax=0.20,
    symbolic x coords={Douban-Movie, Douban-Book},
    xtick=data,
    ytick={0,0.06,0.12,0.18},
    legend style={at={(0.76,0.98)}, anchor=north,legend columns=1, column sep=0.2cm, draw=none, font=\LARGE},
]
\addplot coordinates {
(Douban-Movie,0.1249) (Douban-Book,0.1123)};
\addplot coordinates {
(Douban-Movie,0.1424) (Douban-Book,0.1099)};
\addplot coordinates {
(Douban-Movie,0.1666) (Douban-Book,0.1379)};
\legend{Concatenation, Element-wise sum, Attention}
\end{axis}
\end{tikzpicture}}
\subfigure[]{
\begin{tikzpicture}[scale=0.38]
\pgfplotsset{%
    width=0.60\textwidth,
    height=0.50\textwidth
}
\begin{axis}[
    ybar,
    bar width=15pt,
    ylabel={NGCG@10},
    ylabel style ={font = \LARGE},
    xlabel style ={font = \LARGE},
    enlarge x limits={abs=2.5cm},
    scaled ticks=false,
    tick label style={/pgf/number format/fixed, font=\LARGE},
    ymin=0, ymax=0.12,
    symbolic x coords={Douban-Movie, Douban-Book},
    xtick=data,
    ytick={0,0.03,0.06,0.09,0.12},
    legend style={at={(0.76,0.98)}, anchor=north,legend columns=1, column sep=0.2cm, draw=none, font=\LARGE},
]
\addplot coordinates {
(Douban-Movie,0.0602) (Douban-Book,0.0591)};
\addplot coordinates {
(Douban-Movie,0.0652) (Douban-Book,0.0575)};
\addplot coordinates {
(Douban-Movie,0.0916) (Douban-Book,0.0818)};
\legend{Concatenation, Element-wise sum, Attention}
\end{axis}
\end{tikzpicture}}
\caption{(a)-(b): Performance comparison between our model and its four variants. (c)-(d): Performance comparison of adopted information fusion approaches.}
\end{figure}

\subsection{Impact of Various Components of User Preferences (for RQ3)}
To demonstrate that all three components of user preferences, i.e., domain-share, domain-specific and domain-independent information, are essential and effective for recommendation, and do not require transfer of domain-independent user preferences, we compare DIDA-CDR with its four variants, including DIDA-CDR (w/o sha.), DIDA-CDR (w/o spe.), DIDA-CDR (w/o ind.) and DIDA-CDR (transfer ind.). Fig. 3(a)-(b) shows the performance comparison between our model and the above four variants\footnote{Due to space limitation, we only show the results on Task 1 in Fig. 3, i.e., only the results on the pair of datasets consisting of Douban-Movie and Douban-Book are presented, and similar trends can be observed for results on the other omitted tasks. Similarly, Fig. 4 and Fig. 5 only shows the results on Task 1 for the same reason above.}. The differences between various variants and the impact of each component of user preferences are elaborated in the following subsections.

\subsubsection{\textbf{Impact of Domain-Shared User Preferences}}
\textbf{DIDA-CDR (w/o sha.)} extracts the domain-shared user preferences from both domains, but does not transfer them to any domain. From Fig. 3(a)-(b), we can see that our proposed DIDA-CDR outperforms DIDA-CDR (w/o sha.) with an average improvement of 17.21\%. This is because the domain-shared user preferences are valuable information, which plays an important role in cross-domain recommendation and can improve the recommendation performance on both domains simultaneously.

\begin{figure}[t]
 \setlength{\belowcaptionskip}{-0.12in}
 \centering
 \footnotesize
 \subfigure[]{
  \begin{tikzpicture}
  \begin{axis}[
  width=4.6cm,
  height=3.8cm,
  xmin=0.9, xmax=4.1,
  ymin=0.078, ymax=0.18,
  ylabel={HR@10},
  ylabel style={yshift=-0.4cm},
  xlabel={$l$},
  xlabel style={yshift=0.2cm},
  xtick={1,2,3,4},
  yticklabel style={/pgf/number format/.cd,fixed,precision=3},
  ytick={0,0.09,0.13,0.17},
  scaled ticks=false,
  legend style={at={(0.49,0.34)}, font=\tiny, anchor=north,legend columns=1, draw=none, fill=none,},
  ymajorgrids=true,
  grid style=dashed,
  ]
  \addplot[color=red3,
  mark=*,
  mark options={solid},
  line width=1pt,mark size=1.5pt,
  smooth] coordinates {
   (1,0.1162)
   (2,0.1666)
   (3,0.1616)
   (4,0.1438)
   };
  \addplot[ color=red1,
  mark=square,
  mark options={solid},
  line width=1pt,mark size=1.5pt,
  smooth] coordinates {
   (1,0.1111)
   (2,0.1379)
   (3,0.1235)
   (4,0.1123)
   };
  \legend{Douban-Movie, Douban-Book}
  \end{axis}
  \end{tikzpicture}}
  \hspace{0in}
 \subfigure[]{
  \begin{tikzpicture}
  \begin{axis}[
  width=4.6cm,
  height=3.8cm,
  xmin=0.9, xmax=4.1,
  ymin=0.038, ymax=0.10,
  ylabel={NDCG@10},
  ylabel style={yshift=-0.4cm},
  xlabel={$l$},
  xlabel style={yshift=0.2cm},
  xtick={1,2,3,4},
  yticklabel style={/pgf/number format/.cd,fixed,precision=3},
  ytick={0,0.048,0.068,0.088},
  scaled ticks=false,
  legend style={at={(0.49,0.34)}, font=\tiny, anchor=north,legend columns=1, draw=none, fill=none,},
  ymajorgrids=true,
  grid style=dashed,
  ]
  \addplot [color=red3,
  mark=*,
  mark options={solid},
  line width=1pt,mark size=1.5pt,
  smooth]coordinates {
   (1,0.059)
   (2,0.0916)
   (3,0.0865)
   (4,0.0638)
  };
  \addplot [color=red1,
  mark=square,
  mark options={solid},
  line width=1pt,mark size=1.5pt,
  smooth]coordinates {
   (1,0.0505)
   (2,0.0818)
   (3,0.0762)
   (4,0.0564)
  };
  \legend{Douban-Movie, Douban-Book}
  \end{axis}
  \end{tikzpicture}}
  \hspace{0in}
  \subfigure[]{
  \begin{tikzpicture}
  \begin{axis}[
  width=4.6cm,
  height=3.8cm,
  xmin=0.9, xmax=5.1,
  ymin=0.078, ymax=0.18,
  ylabel={HR@10},
  ylabel style={yshift=-0.4cm},
  xlabel={$\alpha$},
  xlabel style={yshift=0.2cm},
  xtick={1,2,3,4,5},
  xticklabels={0.1,0.5,1,2,5},
  yticklabel style={/pgf/number format/.cd,fixed,precision=3},
  ytick={0,0.09,0.13,0.17},
  scaled ticks=false,
  legend style={at={(0.49,0.34)}, font=\tiny, anchor=north,legend columns=1, draw=none, fill=none,},
  ymajorgrids=true,
  grid style=dashed,
  ]
  \addplot[color=red3,
  mark=*,
  mark options={solid},
  line width=1pt,mark size=1.5pt,
  smooth] coordinates {
   (1,0.1468)
   (2,0.1545)
   (3,0.1666)
   (4,0.1584)
   (5,0.1380)
   };
  \addplot[ color=red1,
  mark=square,
  mark options={solid},
  line width=1pt,mark size=1.5pt,
  smooth] coordinates {
   (1,0.1070)
   (2,0.1123)
   (3,0.1379)
   (4,0.1105)
   (5,0.1059)
   };
  \legend{Douban-Movie, Douban-Book}
  \end{axis}
  \end{tikzpicture}}
  \hspace{0in}
 \subfigure[]{
  \begin{tikzpicture}
  \begin{axis}[
  width=4.6cm,
  height=3.8cm,
  xmin=0.9, xmax=5.1,
  ymin=0.038, ymax=0.10,
  ylabel={NDCG@10},
  ylabel style={yshift=-0.4cm},
  xlabel={$\alpha$},
  xlabel style={yshift=0.2cm},
  xtick={1,2,3,4,5},
  xticklabels={0.1,0.5,1,2,5},
  yticklabel style={/pgf/number format/.cd,fixed,precision=3},
  ytick={0,0.048,0.068,0.088},
  scaled ticks=false,
  legend style={at={(0.49,0.34)}, font=\tiny, anchor=north,legend columns=1, draw=none, fill=none,},
  ymajorgrids=true,
  grid style=dashed,
  ]
  \addplot [color=red3,
  mark=*,
  mark options={solid},
  line width=1pt,mark size=1.5pt,
  smooth]coordinates {
   (1,0.0757)
   (2,0.0859)
   (3,0.0916)
   (4,0.0836)
   (5,0.0604)
  };
  \addplot [color=red1,
  mark=square,
  mark options={solid},
  line width=1pt,mark size=1.5pt,
  smooth]coordinates {
   (1,0.0560)
   (2,0.0587)
   (3,0.0818)
   (4,0.0586)
   (5,0.0580)
  };
  \legend{Douban-Movie, Douban-Book}
  \end{axis}
  \end{tikzpicture}}
 \caption{(a)-(b): Impact of the number of GCN layers. (c)-(d): Impact of $\alpha$.}
\end{figure}
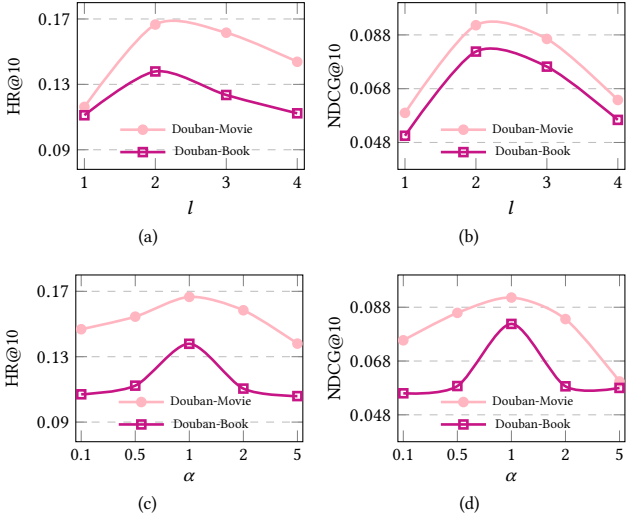

\subsubsection{\textbf{Impact of Domain-Specific User Preferences}}
\textbf{DIDA-CDR (w/o spe.)} does not consider the domain-specific user preferences when making recommendations. From Fig. 3(a)-(b), we can observe that our proposed DIDA-CDR outperforms DIDA-CDR (w/o spe.) with an average improvement of 36.12\%. This is because the domain-specific user preferences are inherent personalized preferences of users in each domain. If it is not considered when making recommendations, the recommendation performance of our model will be significantly reduced.

\subsubsection{\textbf{Impact of Domain-Independent User Preferences}}
\textbf{DIDA-CDR (w/o ind.)} includes the domain-shared and domain-specific user preferences, but does not include the domain-independent user preferences when capturing comprehensive user preferences; \textbf{DIDA-CDR (transfer ind.)} disentangles the domain-independent user preferences and then transfer them to another domain. It can be seen from Fig. 3(a)-(b) that our proposed DIDA-CDR outperforms DIDA-CDR (w/o ind.) and DIDA-CDR (transfer ind.) with an average improvement of 12.40\% and 26.04\%, respectively. In view of this, we make the following qualitative analysis. The domain-independent user preferences seemingly exist in each domain, but actually have different meanings. If they are not included when capturing user preferences, the captured user preferences are incomplete, resulting in the degraded recommendation performance of the model. Moreover, if the domain-independent information is transferred to other domains, it will provide the useless information and cause the performance degradation.

\begin{figure}[t]
 \setlength{\belowcaptionskip}{-0.12in}
 \centering
 \footnotesize
 \subfigure[]{
  \begin{tikzpicture}
  \begin{axis}[
  width=4.5cm,
  height=3.8cm,
  xmin=0.8, xmax=8.2,
  ymin=0.10, ymax=0.178,
  ylabel={HR@10},
  ylabel style={yshift=-0.4cm},
  xlabel={${\mu _1}$},
  xlabel style={yshift=0.2cm},
  xtick={1,2,3,4,5,6,7,8},
  xticklabels={0.1,0.3,0.5,0.7,1,3,5,10},
  yticklabel style={/pgf/number format/.cd,fixed,precision=3},
  ytick={0.11,0.13, 0.15, 0.17},
  scaled ticks=false,
  legend style={at={(0.49,0.34)}, font=\tiny, anchor=north,legend columns=1, draw=none, fill=none,},
  ymajorgrids=true,
  grid style=dashed,
  ]
  \addplot[color=red3,
  mark=*,
  mark options={solid},
  line width=1pt,mark size=1.5pt,
  smooth] coordinates {
   (1,0.1411)
   (2,0.1488)
   (3,0.1533)
   (4,0.1654)
   (5,0.1666)
   (6,0.1642)
   (7,0.1587)
   (8,0.1489)
   };
  \addplot[ color=red1,
  mark=square,
  mark options={solid},
  line width=1pt,mark size=1.5pt,
  smooth] coordinates {
   (1,0.1186)
   (2,0.1250)
   (3,0.1282)
   (4,0.1366)
   (5,0.1379)
   (6,0.1337)
   (7,0.1295)
   (8,0.1258)
   };
  \legend{Douban-Movie, Douban-Book}
  \end{axis}
  \end{tikzpicture}}
  \hspace{0in}
 \subfigure[]{
  \begin{tikzpicture}
  \begin{axis}[
  width=4.5cm,
  height=3.8cm,
  xmin=0.8, xmax=8.2,
  ymin=0.05, ymax=0.108,
  ylabel={NDCG@10},
  ylabel style={yshift=-0.4cm},
  xlabel={${\mu _1}$},
  xlabel style={yshift=0.2cm},
  xtick={1,2,3,4,5,6,7,8},
  xticklabels={0.1,0.3,0.5,0.7,1,3,5,10},
  yticklabel style={/pgf/number format/.cd,fixed,precision=3},
  ytick={0.055,0.07, 0.085, 0.10},
  scaled ticks=false,
  legend style={at={(0.49,0.34)}, font=\tiny, anchor=north,legend columns=1, draw=none, fill=none,},
  ymajorgrids=true,
  grid style=dashed,
  ]
  \addplot [color=red3,
  mark=*,
  mark options={solid},
  line width=1pt,mark size=1.5pt,
  smooth]coordinates {
   (1,0.0763)
   (2,0.0794)
   (3,0.0835)
   (4,0.0907)
   (5,0.0916)
   (6,0.0904)
   (7,0.0863)
   (8,0.0796)
  };
  \addplot [color=red1,
  mark=square,
  mark options={solid},
  line width=1pt,mark size=1.5pt,
  smooth]coordinates {
   (1,0.0627)
   (2,0.0736)
   (3,0.0764)
   (4,0.0810)
   (5,0.0818)
   (6,0.0803)
   (7,0.0779)
   (8,0.0741)
  };
  \legend{Douban-Movie, Douban-Book}
  \end{axis}
  \end{tikzpicture}}
  \hspace{0in}
  \subfigure[]{
  \begin{tikzpicture}
  \begin{axis}[
  width=4.5cm,
  height=3.8cm,
  xmin=0.8, xmax=8.2,
  ymin=0.10, ymax=0.178,
  ylabel={HR@10},
  ylabel style={yshift=-0.4cm},
  xlabel={${\mu _2}$},
  xlabel style={yshift=0.2cm},
  xtick={1,2,3,4,5,6,7,8},
  xticklabels={0.1,0.3,0.5,0.7,1,3,5,10},
  yticklabel style={/pgf/number format/.cd,fixed,precision=3},
  ytick={0.11,0.13, 0.15, 0.17},
  scaled ticks=false,
  legend style={at={(0.49,0.34)}, font=\tiny, anchor=north,legend columns=1, draw=none, fill=none,},
  ymajorgrids=true,
  grid style=dashed,
  ]
  \addplot[color=red3,
  mark=*,
  mark options={solid},
  line width=1pt,mark size=1.5pt,
  smooth] coordinates {
   (1,0.1403)
   (2,0.1474)
   (3,0.1527)
   (4,0.1639)
   (5,0.1666)
   (6,0.1621)
   (7,0.1553)
   (8,0.1480)
   };
  \addplot[ color=red1,
  mark=square,
  mark options={solid},
  line width=1pt,mark size=1.5pt,
  smooth] coordinates {
   (1,0.1181)
   (2,0.1241)
   (3,0.1275)
   (4,0.1328)
   (5,0.1379)
   (6,0.1316)
   (7,0.1284)
   (8,0.1245)
   };
  \legend{Douban-Movie, Douban-Book}
  \end{axis}
  \end{tikzpicture}}
  \hspace{0in}
 \subfigure[]{
  \begin{tikzpicture}
  \begin{axis}[
  width=4.5cm,
  height=3.8cm,
  xmin=0.8, xmax=8.2,
  ymin=0.05, ymax=0.108,
  ylabel={NDCG@10},
  ylabel style={yshift=-0.4cm},
  xlabel={${\mu _2}$},
  xlabel style={yshift=0.2cm},
  xtick={1,2,3,4,5,6,7,8},
  xticklabels={0.1,0.3,0.5,0.7,1,3,5,10},
  yticklabel style={/pgf/number format/.cd,fixed,precision=3},
  ytick={0.055,0.07, 0.085, 0.10},
  scaled ticks=false,
  legend style={at={(0.49,0.34)}, font=\tiny, anchor=north,legend columns=1, draw=none, fill=none,},
  ymajorgrids=true,
  grid style=dashed,
  ]
  \addplot [color=red3,
  mark=*,
  mark options={solid},
  line width=1pt,mark size=1.5pt,
  smooth]coordinates {
   (1,0.0759)
   (2,0.0790)
   (3,0.0826)
   (4,0.0903)
   (5,0.0916)
   (6,0.0899)
   (7,0.0838)
   (8,0.0791)
  };
  \addplot [color=red1,
  mark=square,
  mark options={solid},
  line width=1pt,mark size=1.5pt,
  smooth]coordinates {
   (1,0.0602)
   (2,0.0724)
   (3,0.0758)
   (4,0.0801)
   (5,0.0818)
   (6,0.0773)
   (7,0.0762)
   (8,0.0729)
  };
  \legend{Douban-Movie, Douban-Book}
  \end{axis}
  \end{tikzpicture}}
 \caption{(a)-(b): Impact of ${\mu _1}$. (c)-(d): Impact of ${\mu _2}$.}
\end{figure}
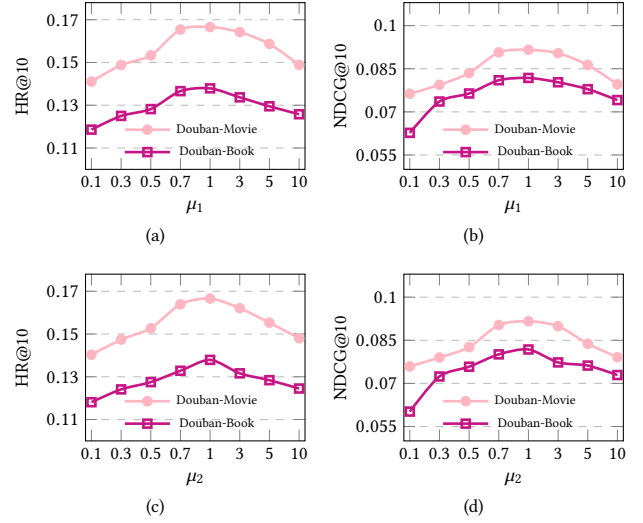

\subsection{Impact of Various Information Fusion Approaches (for RQ4)}
After we obtain three essential user preference components, we compare various information fusion approaches, i.e., concatenation, element-wise sum, and attention, to fuse them into comprehensive user preferences. The performance comparison of three used information fusion approaches is presented in Fig. 3(c)-(d). We find that when our model utilizes the attention mechanism for information fusion, it improves the variants using concatenation and summation by an average of 28.37\% and 20.69\%, respectively. This is because the attention mechanism can not only capture the relationship between various components, but also selectively highlight the key information and weaken the redundant information by learning weights. In this paper, the attention mechanism measures the importance of domain-shared, domain-specific and domain-independent information to comprehensive user preferences through weights, and weakens the redundant information between the domain-shared and domain-independent user preferences, thus enabling our model to achieve better recommendation results.

\subsection{Parameter Sensitivity (for RQ5)}
\subsubsection{\textbf{Impact of $l$.}} Stacking too many layers when training a deep GCN is prone to over-smoothing \cite{li2018deeper}. In order to explore this effect, we investigate the number of GCN layers $l$ in the range of $\{1, 2, 3, 4\}$ in the experiments and report the experimental results in Fig. 4(a)-(b). It can be observed that 2-layer GCN is significantly better than 1-layer GCN, which indicates that stacking a moderate number of layers is beneficial for mining higher-order user-item relationships. However, the recommendation performance of our proposed DIDA-CDR drops on some datasets when $l = 3$ and degrades even more when $l = 4$. The reason may be that when the number of layers is greater than 3, the problem of over-smoothing occurs, resulting in the fact that increasing the number of layers at this time will in turn reduce the recommendation performance of our model.

\subsubsection{\textbf{Impact of $\alpha$.}} $Beta(\alpha ,\alpha )$ is Uniform distribution when $\alpha  = 1$, Bimodal distribution when $\alpha < 1$ and Bell-shaped distribution when $\alpha > 1$ \cite{chen2022graph}. In order to explore from which distribution sampling $\lambda $ can help the recommendation performance of model the most, we search $\alpha$ in the range of $\{0.1, 0.5, 1, 2, 5\}$. The performance comparison is illustrated in Fig. 4(c)-(d). We can see that the best performance of our proposed DIDA-CDR is achieved when $\alpha  = 1$, which presents that $\lambda $ sampled from the Uniform distribution can effectively mix the user embeddings in the richer domain and the sparser domain, thereby effectively alleviating the data imbalance and improving the recommendation performance of model. In contrast, when $\alpha < 1$ or $\alpha > 1$, sampling $\lambda $ from $Beta(\alpha ,\alpha )$ will result in the performance degradation.

\subsubsection{\textbf{Impact of ${\mu _1}$ and ${\mu _2}$.}} To explore the effect of weights of domain classification losses on the overall recommendation performance of our DIDA-CDR, we vary ${\mu _1}$ and ${\mu _2}$ in $\{0.1, 0.3, 0.5, 0.7, 1, 3, 5, 10\}$. The results are reported in Fig. 5. It can be seen from Fig. 5(a)-(b) that the contribution of domain classification loss ${\mathcal L_{cl{s_1}}}$ to decoupling more accurate domain-specific user preferences is small when ${\mu _1} \rightarrow 0$. When ${\mu _1}$ is too large, the domain classification loss ${\mathcal L_{cl{s_1}}}$ receives more attention during the learning process. As a result, the contribution of prediction loss ${\mathcal L_{prd}}$ to the model is weakened, which reduces the recommendation performance of the model. Similarly, from Fig. 5(c)-(d), a similar trend can be observed for the weights ${\mu _2}$ of domain classification loss $L_{cl{s_2}}$. Empirically, we choose ${\mu _1} = {\mu _2} = 1$.

\section{Conclusion and Future Work}
In this paper, we have proposed a novel Disentanglement-based framework with Interpolative Data Augmentation for dual-target Cross-Domain Recommendation, called DIDA-CDR. DIDA-CDR consists of an interpolative data augmentation approach to generating both relevant and diverse augmented user representations to augment the sparser domain and explore the potential user preferences, and a user preference disentanglement module to decouple essential components of user preferences to capture comprehensive user preferences, all of which help improve the recommendation performance on both domains simultaneously. Also, we have conducted extensive experiments on five real-world datasets to show the significant superiority of DIDA-CDR over the state-of-the-art methods. In the future, we plan to extend our model to multi-target CDR scenarios where users partially overlap.

\begin{acks}
This work is partially supported by ARC Discovery Projects DP200101441 and DP230100676.
\end{acks}

\balance

\bibliographystyle{ACM-Reference-Format}
\bibliography{DIDA-CDR}


\begin{thebibliography}{62}


\ifx \showCODEN    \undefined \def \showCODEN     #1{\unskip}     \fi
\ifx \showDOI      \undefined \def \showDOI       #1{#1}\fi
\ifx \showISBNx    \undefined \def \showISBNx     #1{\unskip}     \fi
\ifx \showISBNxiii \undefined \def \showISBNxiii  #1{\unskip}     \fi
\ifx \showISSN     \undefined \def \showISSN      #1{\unskip}     \fi
\ifx \showLCCN     \undefined \def \showLCCN      #1{\unskip}     \fi
\ifx \shownote     \undefined \def \shownote      #1{#1}          \fi
\ifx \showarticletitle \undefined \def \showarticletitle #1{#1}   \fi
\ifx \showURL      \undefined \def \showURL       {\relax}        \fi
\providecommand\bibfield[2]{#2}
\providecommand\bibinfo[2]{#2}
\providecommand\natexlab[1]{#1}
\providecommand\showeprint[2][]{arXiv:#2}

\bibitem[Bengio et~al\mbox{.}(2013)]%
        {bengio2013representation}
\bibfield{author}{\bibinfo{person}{Yoshua Bengio}, \bibinfo{person}{Aaron
  Courville}, {and} \bibinfo{person}{Pascal Vincent}.}
  \bibinfo{year}{2013}\natexlab{}.
\newblock \showarticletitle{Representation Learning: A Review and New
  Perspectives}.
\newblock \bibinfo{journal}{\emph{TPAMI}} (\bibinfo{year}{2013}),
  \bibinfo{pages}{1798--1828}.
\newblock


\bibitem[Bian et~al\mbox{.}(2022)]%
        {bian2022relevant}
\bibfield{author}{\bibinfo{person}{Shuqing Bian}, \bibinfo{person}{Wayne~Xin
  Zhao}, \bibinfo{person}{Jinpeng Wang}, {and} \bibinfo{person}{Ji-Rong Wen}.}
  \bibinfo{year}{2022}\natexlab{}.
\newblock \showarticletitle{A Relevant and Diverse Retrieval-enhanced Data
  Augmentation Framework for Sequential Recommendation}. In
  \bibinfo{booktitle}{\emph{CIKM}}. \bibinfo{pages}{2923--2932}.
\newblock


\bibitem[Cao et~al\mbox{.}(2022a)]%
        {cao2022disencdr}
\bibfield{author}{\bibinfo{person}{Jiangxia Cao}, \bibinfo{person}{Xixun Lin},
  \bibinfo{person}{Xin Cong}, \bibinfo{person}{Jing Ya},
  \bibinfo{person}{Tingwen Liu}, {and} \bibinfo{person}{Bin Wang}.}
  \bibinfo{year}{2022}\natexlab{a}.
\newblock \showarticletitle{DisenCDR: Learning Disentangled Representations for
  Cross-Domain Recommendation}. In \bibinfo{booktitle}{\emph{SIGIR}}.
  \bibinfo{pages}{267--277}.
\newblock


\bibitem[Cao et~al\mbox{.}(2022b)]%
        {cao2022cross}
\bibfield{author}{\bibinfo{person}{Jiangxia Cao}, \bibinfo{person}{Jiawei
  Sheng}, \bibinfo{person}{Xin Cong}, \bibinfo{person}{Tingwen Liu}, {and}
  \bibinfo{person}{Bin Wang}.} \bibinfo{year}{2022}\natexlab{b}.
\newblock \showarticletitle{Cross-Domain Recommendation to Cold-Start Users via
  Variational Information Bottleneck}. In \bibinfo{booktitle}{\emph{ICDE}}.
  \bibinfo{pages}{2209--2223}.
\newblock


\bibitem[Chen et~al\mbox{.}(2022)]%
        {chen2022graph}
\bibfield{author}{\bibinfo{person}{Huiyuan Chen},
  \bibinfo{person}{Chin-Chia~Michael Yeh}, \bibinfo{person}{Fei Wang}, {and}
  \bibinfo{person}{Hao Yang}.} \bibinfo{year}{2022}\natexlab{}.
\newblock \showarticletitle{Graph Neural Transport Networks with Non-local
  Attentions for Recommender Systems}. In \bibinfo{booktitle}{\emph{WWW}}.
  \bibinfo{pages}{1955--1964}.
\newblock


\bibitem[Chen et~al\mbox{.}(2018)]%
        {chen2018isolating}
\bibfield{author}{\bibinfo{person}{Ricky T.~Q. Chen}, \bibinfo{person}{Xuechen
  Li}, \bibinfo{person}{Roger~B Grosse}, {and} \bibinfo{person}{David~K
  Duvenaud}.} \bibinfo{year}{2018}\natexlab{}.
\newblock \showarticletitle{Isolating Sources of Disentanglement in Variational
  Autoencoders}. In \bibinfo{booktitle}{\emph{NeurIPS}}.
\newblock


\bibitem[Choi et~al\mbox{.}(2022)]%
        {choi2022based}
\bibfield{author}{\bibinfo{person}{Yoonhyuk Choi}, \bibinfo{person}{Jiho Choi},
  \bibinfo{person}{Taewook Ko}, \bibinfo{person}{Hyungho Byun}, {and}
  \bibinfo{person}{Chong-Kwon Kim}.} \bibinfo{year}{2022}\natexlab{}.
\newblock \showarticletitle{Review-Based Domain Disentanglement without
  Duplicate Users or Contexts for Cross-Domain Recommendation}. In
  \bibinfo{booktitle}{\emph{CIKM}}. \bibinfo{pages}{293--303}.
\newblock


\bibitem[Cui et~al\mbox{.}(2020)]%
        {cui2020herograph}
\bibfield{author}{\bibinfo{person}{Qiang Cui}, \bibinfo{person}{Tao Wei},
  \bibinfo{person}{Yafeng Zhang}, {and} \bibinfo{person}{Qing Zhang}.}
  \bibinfo{year}{2020}\natexlab{}.
\newblock \showarticletitle{HeroGRAPH: A Heterogeneous Graph Framework for
  Multi-Target Cross-Domain Recommendation}. In
  \bibinfo{booktitle}{\emph{RecSys}}.
\newblock


\bibitem[Fu et~al\mbox{.}(2019)]%
        {fu2019deeply}
\bibfield{author}{\bibinfo{person}{Wenjing Fu}, \bibinfo{person}{Zhaohui Peng},
  \bibinfo{person}{Senzhang Wang}, \bibinfo{person}{Yang Xu}, {and}
  \bibinfo{person}{Jin Li}.} \bibinfo{year}{2019}\natexlab{}.
\newblock \showarticletitle{Deeply Fusing Reviews and Contents for Cold Start
  Users in Cross-Domain Recommendation Systems}. In
  \bibinfo{booktitle}{\emph{AAAI}}. \bibinfo{pages}{94--101}.
\newblock


\bibitem[Fu et~al\mbox{.}(2022)]%
        {fu2022generalized}
\bibfield{author}{\bibinfo{person}{Yuqian Fu}, \bibinfo{person}{Yanwei Fu},
  \bibinfo{person}{Jingjing Chen}, {and} \bibinfo{person}{Yu-Gang Jiang}.}
  \bibinfo{year}{2022}\natexlab{}.
\newblock \showarticletitle{Generalized Meta-FDMixup: Cross-Domain Few-Shot
  Learning Guided by Labeled Target Data}.
\newblock \bibinfo{journal}{\emph{TIP}} (\bibinfo{year}{2022}),
  \bibinfo{pages}{7078--7090}.
\newblock


\bibitem[Fu et~al\mbox{.}(2021)]%
        {fu2021meta}
\bibfield{author}{\bibinfo{person}{Yuqian Fu}, \bibinfo{person}{Yanwei Fu},
  {and} \bibinfo{person}{Yu-Gang Jiang}.} \bibinfo{year}{2021}\natexlab{}.
\newblock \showarticletitle{Meta-FDMixup: Cross-Domain Few-Shot Learning Guided
  by Labeled Target Data}. In \bibinfo{booktitle}{\emph{MM}}.
  \bibinfo{pages}{5326--5334}.
\newblock


\bibitem[Gibbs(2015)]%
        {gibbs2015writing}
\bibfield{author}{\bibinfo{person}{Anna Gibbs}.}
  \bibinfo{year}{2015}\natexlab{}.
\newblock \showarticletitle{Writing as method: attunement, resonance, and
  rhythm}.
\newblock \bibinfo{journal}{\emph{Affective methodologies: Developing cultural
  research strategies for the study of affect}} (\bibinfo{year}{2015}),
  \bibinfo{pages}{222--236}.
\newblock


\bibitem[Gonzalez-Garcia et~al\mbox{.}(2018)]%
        {gonzalez2018image}
\bibfield{author}{\bibinfo{person}{Abel Gonzalez-Garcia},
  \bibinfo{person}{Joost van~de Weijer}, {and} \bibinfo{person}{Yoshua
  Bengio}.} \bibinfo{year}{2018}\natexlab{}.
\newblock \showarticletitle{Image-to-image Translation for Cross-domain
  Disentanglement}. In \bibinfo{booktitle}{\emph{NeurIPS}}.
\newblock


\bibitem[Guo et~al\mbox{.}(2023)]%
        {guo2023disentangled}
\bibfield{author}{\bibinfo{person}{Xiaobo Guo}, \bibinfo{person}{Shaoshuai Li},
  \bibinfo{person}{Naicheng Guo}, \bibinfo{person}{Jiangxia Cao},
  \bibinfo{person}{Xiaolei Liu}, \bibinfo{person}{Qiongxu Ma},
  \bibinfo{person}{Runsheng Gan}, {and} \bibinfo{person}{Yunan Zhao}.}
  \bibinfo{year}{2023}\natexlab{}.
\newblock \showarticletitle{Disentangled Representations Learning for
  Multi-target Cross-domain Recommendation}.
\newblock \bibinfo{journal}{\emph{TOIS}} (\bibinfo{year}{2023}),
  \bibinfo{pages}{1--27}.
\newblock


\bibitem[Hauser and McDermott(2003)]%
        {hauser2003evolution}
\bibfield{author}{\bibinfo{person}{Marc~D Hauser} {and} \bibinfo{person}{Josh
  McDermott}.} \bibinfo{year}{2003}\natexlab{}.
\newblock \showarticletitle{The evolution of the music faculty: A comparative
  perspective}.
\newblock \bibinfo{journal}{\emph{Nature neuroscience}} (\bibinfo{year}{2003}),
  \bibinfo{pages}{663--668}.
\newblock


\bibitem[He et~al\mbox{.}(2020)]%
        {he2020lightgcn}
\bibfield{author}{\bibinfo{person}{Xiangnan He}, \bibinfo{person}{Kuan Deng},
  \bibinfo{person}{Xiang Wang}, \bibinfo{person}{Yan Li},
  \bibinfo{person}{Yongdong Zhang}, {and} \bibinfo{person}{Meng Wang}.}
  \bibinfo{year}{2020}\natexlab{}.
\newblock \showarticletitle{Lightgcn: Simplifying and Powering Graph
  Convolution Network for Recommendation}. In
  \bibinfo{booktitle}{\emph{SIGIR}}. \bibinfo{pages}{639--648}.
\newblock


\bibitem[Hou et~al\mbox{.}(2022)]%
        {hou2022towards}
\bibfield{author}{\bibinfo{person}{Yupeng Hou}, \bibinfo{person}{Shanlei Mu},
  \bibinfo{person}{Wayne~Xin Zhao}, \bibinfo{person}{Yaliang Li},
  \bibinfo{person}{Bolin Ding}, {and} \bibinfo{person}{Ji-Rong Wen}.}
  \bibinfo{year}{2022}\natexlab{}.
\newblock \showarticletitle{Towards Universal Sequence Representation Learning
  for Recommender Systems}. In \bibinfo{booktitle}{\emph{KDD}}.
  \bibinfo{pages}{585--593}.
\newblock


\bibitem[Hu et~al\mbox{.}(2018)]%
        {hu2018conet}
\bibfield{author}{\bibinfo{person}{Guangneng Hu}, \bibinfo{person}{Yu Zhang},
  {and} \bibinfo{person}{Qiang Yang}.} \bibinfo{year}{2018}\natexlab{}.
\newblock \showarticletitle{Conet: Collaborative Cross Networks for
  Cross-Domain Recommendation}. In \bibinfo{booktitle}{\emph{CIKM}}.
  \bibinfo{pages}{667--676}.
\newblock


\bibitem[Hu et~al\mbox{.}(2019)]%
        {hu2019transfer}
\bibfield{author}{\bibinfo{person}{Guangneng Hu}, \bibinfo{person}{Yu Zhang},
  {and} \bibinfo{person}{Qiang Yang}.} \bibinfo{year}{2019}\natexlab{}.
\newblock \showarticletitle{Transfer Meets Hybrid: A Synthetic Approach for
  Cross-Domain Collaborative Filtering with Text}. In
  \bibinfo{booktitle}{\emph{WWW}}. \bibinfo{pages}{2822--2829}.
\newblock


\bibitem[Huang et~al\mbox{.}(2021)]%
        {huang2021mixgcf}
\bibfield{author}{\bibinfo{person}{Tinglin Huang}, \bibinfo{person}{Yuxiao
  Dong}, \bibinfo{person}{Ming Ding}, \bibinfo{person}{Zhen Yang},
  \bibinfo{person}{Wenzheng Feng}, \bibinfo{person}{Xinyu Wang}, {and}
  \bibinfo{person}{Jie Tang}.} \bibinfo{year}{2021}\natexlab{}.
\newblock \showarticletitle{MixGCF: An Improved Training Method for Graph
  Neural Network-based Recommender Systems}. In
  \bibinfo{booktitle}{\emph{KDD}}. \bibinfo{pages}{665--674}.
\newblock


\bibitem[Kanagawa et~al\mbox{.}(2019)]%
        {kanagawa2019cross}
\bibfield{author}{\bibinfo{person}{Heishiro Kanagawa}, \bibinfo{person}{Hayato
  Kobayashi}, \bibinfo{person}{Nobuyuki Shimizu}, \bibinfo{person}{Yukihiro
  Tagami}, {and} \bibinfo{person}{Taiji Suzuki}.}
  \bibinfo{year}{2019}\natexlab{}.
\newblock \showarticletitle{Cross-domain Recommendation via Deep Domain
  Adaptation}. In \bibinfo{booktitle}{\emph{ECIR}}. \bibinfo{pages}{20--29}.
\newblock


\bibitem[Kingma and Ba(2015)]%
        {kingma2014adam}
\bibfield{author}{\bibinfo{person}{Diederik~P Kingma} {and}
  \bibinfo{person}{Jimmy Ba}.} \bibinfo{year}{2015}\natexlab{}.
\newblock \showarticletitle{Adam: A Method for Stochastic Optimization}. In
  \bibinfo{booktitle}{\emph{ICLR}}.
\newblock


\bibitem[Kipf and Welling(2017)]%
        {KipfW17}
\bibfield{author}{\bibinfo{person}{Thomas~N. Kipf} {and} \bibinfo{person}{Max
  Welling}.} \bibinfo{year}{2017}\natexlab{}.
\newblock \showarticletitle{Semi-Supervised Classification with Graph
  Convolutional Networks}. In \bibinfo{booktitle}{\emph{ICLR}}.
\newblock


\bibitem[Krichene and Rendle(2022)]%
        {krichene2022sampled}
\bibfield{author}{\bibinfo{person}{Walid Krichene} {and}
  \bibinfo{person}{Steffen Rendle}.} \bibinfo{year}{2022}\natexlab{}.
\newblock \showarticletitle{On Sampled Metrics for Item Recommendation}.
\newblock \bibinfo{journal}{\emph{CACM}} (\bibinfo{year}{2022}),
  \bibinfo{pages}{75--83}.
\newblock


\bibitem[Li et~al\mbox{.}(2023)]%
        {li2023one}
\bibfield{author}{\bibinfo{person}{Chenglin Li}, \bibinfo{person}{Yuanzhen
  Xie}, \bibinfo{person}{Chenyun Yu}, \bibinfo{person}{Bo Hu},
  \bibinfo{person}{Zang Li}, \bibinfo{person}{Guoqiang Shu},
  \bibinfo{person}{Xiaohu Qie}, {and} \bibinfo{person}{Di Niu}.}
  \bibinfo{year}{2023}\natexlab{}.
\newblock \showarticletitle{One for All, All for One: Learning and Transferring
  User Embeddings for Cross-Domain Recommendation}. In
  \bibinfo{booktitle}{\emph{WSDM}}. \bibinfo{pages}{366--374}.
\newblock


\bibitem[Li and Tuzhilin(2020)]%
        {li2020ddtcdr}
\bibfield{author}{\bibinfo{person}{Pan Li} {and} \bibinfo{person}{Alexander
  Tuzhilin}.} \bibinfo{year}{2020}\natexlab{}.
\newblock \showarticletitle{Ddtcdr: Deep Dual Transfer Cross Domain
  Recommendation}. In \bibinfo{booktitle}{\emph{WSDM}}.
  \bibinfo{pages}{331--339}.
\newblock


\bibitem[Li et~al\mbox{.}(2018)]%
        {li2018deeper}
\bibfield{author}{\bibinfo{person}{Qimai Li}, \bibinfo{person}{Zhichao Han},
  {and} \bibinfo{person}{Xiao-Ming Wu}.} \bibinfo{year}{2018}\natexlab{}.
\newblock \showarticletitle{Deeper Insights into Graph Convolutional Networks
  for Semi-supervised Learning}. In \bibinfo{booktitle}{\emph{AAAI}}.
  \bibinfo{pages}{3538--3545}.
\newblock


\bibitem[Liu et~al\mbox{.}(2020b)]%
        {liu2020exploiting}
\bibfield{author}{\bibinfo{person}{Jian Liu}, \bibinfo{person}{Pengpeng Zhao},
  \bibinfo{person}{Fuzhen Zhuang}, \bibinfo{person}{Yanchi Liu},
  \bibinfo{person}{Victor~S Sheng}, \bibinfo{person}{Jiajie Xu},
  \bibinfo{person}{Xiaofang Zhou}, {and} \bibinfo{person}{Hui Xiong}.}
  \bibinfo{year}{2020}\natexlab{b}.
\newblock \showarticletitle{Exploiting Aesthetic Preference in Deep Cross
  Networks for Cross-Domain Recommendation}. In
  \bibinfo{booktitle}{\emph{WWW}}. \bibinfo{pages}{2768--2774}.
\newblock


\bibitem[Liu et~al\mbox{.}(2020a)]%
        {liu2020cross}
\bibfield{author}{\bibinfo{person}{Meng Liu}, \bibinfo{person}{Jianjun Li},
  \bibinfo{person}{Guohui Li}, {and} \bibinfo{person}{Peng Pan}.}
  \bibinfo{year}{2020}\natexlab{a}.
\newblock \showarticletitle{Cross Domain Recommendation via Bi-directional
  Transfer Graph Collaborative Filtering Networks}. In
  \bibinfo{booktitle}{\emph{CIKM}}. \bibinfo{pages}{885--894}.
\newblock


\bibitem[Liu et~al\mbox{.}(2022)]%
        {liu2022exploiting}
\bibfield{author}{\bibinfo{person}{Weiming Liu}, \bibinfo{person}{Xiaolin
  Zheng}, \bibinfo{person}{Jiajie Su}, \bibinfo{person}{Mengling Hu},
  \bibinfo{person}{Yanchao Tan}, {and} \bibinfo{person}{Chaochao Chen}.}
  \bibinfo{year}{2022}\natexlab{}.
\newblock \showarticletitle{Exploiting Variational Domain-Invariant User
  Embedding for Partially Overlapped Cross Domain Recommendation}. In
  \bibinfo{booktitle}{\emph{SIGIR}}. \bibinfo{pages}{312--321}.
\newblock


\bibitem[Liu et~al\mbox{.}(2019)]%
        {liu2019jscn}
\bibfield{author}{\bibinfo{person}{Zhiwei Liu}, \bibinfo{person}{Lei Zheng},
  \bibinfo{person}{Jiawei Zhang}, \bibinfo{person}{Jiayu Han}, {and}
  \bibinfo{person}{S~Yu Philip}.} \bibinfo{year}{2019}\natexlab{}.
\newblock \showarticletitle{JSCN: Joint Spectral Convolutional Network for
  Cross Domain Recommendation}. In \bibinfo{booktitle}{\emph{Big Data}}.
  \bibinfo{pages}{850--859}.
\newblock


\bibitem[Loni et~al\mbox{.}(2014)]%
        {loni2014cross}
\bibfield{author}{\bibinfo{person}{Babak Loni}, \bibinfo{person}{Yue Shi},
  \bibinfo{person}{Martha Larson}, {and} \bibinfo{person}{Alan Hanjalic}.}
  \bibinfo{year}{2014}\natexlab{}.
\newblock \showarticletitle{Cross-domain Collaborative Filtering with
  Factorization Machines}. In \bibinfo{booktitle}{\emph{ECIR}}.
  \bibinfo{pages}{656--661}.
\newblock


\bibitem[Ma et~al\mbox{.}(2019)]%
        {ma2019learning}
\bibfield{author}{\bibinfo{person}{Jianxin Ma}, \bibinfo{person}{Chang Zhou},
  \bibinfo{person}{Peng Cui}, \bibinfo{person}{Hongxia Yang}, {and}
  \bibinfo{person}{Wenwu Zhu}.} \bibinfo{year}{2019}\natexlab{}.
\newblock \showarticletitle{Learning Disentangled Representations for
  Recommendation}. In \bibinfo{booktitle}{\emph{NeurIPS}}.
  \bibinfo{pages}{5711--5722}.
\newblock


\bibitem[Ma et~al\mbox{.}(2020)]%
        {ma2020disentangled}
\bibfield{author}{\bibinfo{person}{Jianxin Ma}, \bibinfo{person}{Chang Zhou},
  \bibinfo{person}{Hongxia Yang}, \bibinfo{person}{Peng Cui},
  \bibinfo{person}{Xin Wang}, {and} \bibinfo{person}{Wenwu Zhu}.}
  \bibinfo{year}{2020}\natexlab{}.
\newblock \showarticletitle{Disentangled Self-Supervision in Sequential
  Recommenders}. In \bibinfo{booktitle}{\emph{KDD}}. \bibinfo{pages}{483--491}.
\newblock


\bibitem[Man et~al\mbox{.}(2017)]%
        {man2017cross}
\bibfield{author}{\bibinfo{person}{Tong Man}, \bibinfo{person}{Huawei Shen},
  \bibinfo{person}{Xiaolong Jin}, {and} \bibinfo{person}{Xueqi Cheng}.}
  \bibinfo{year}{2017}\natexlab{}.
\newblock \showarticletitle{Cross-Domain Recommendation: An Embedding and
  Mapping Approach}. In \bibinfo{booktitle}{\emph{IJCAI}}.
  \bibinfo{pages}{2464--2470}.
\newblock


\bibitem[Meng et~al\mbox{.}(2021)]%
        {meng2021graph}
\bibfield{author}{\bibinfo{person}{Zaiqiao Meng}, \bibinfo{person}{Siwei Liu},
  \bibinfo{person}{Craig Macdonald}, {and} \bibinfo{person}{Iadh Ounis}.}
  \bibinfo{year}{2021}\natexlab{}.
\newblock \showarticletitle{Graph Neural Pre-training for Enhancing
  Recommendations using Side Information}.
\newblock \bibinfo{journal}{\emph{arXiv preprint arXiv:2107.03936}}
  (\bibinfo{year}{2021}).
\newblock


\bibitem[Sahu and Dwivedi(2020)]%
        {sahu2020knowledge}
\bibfield{author}{\bibinfo{person}{Ashish~Kumar Sahu} {and}
  \bibinfo{person}{Pragya Dwivedi}.} \bibinfo{year}{2020}\natexlab{}.
\newblock \showarticletitle{Knowledge Transfer by Domain-independent User
  Latent Factor for Cross-domain Recommender Systems}.
\newblock \bibinfo{journal}{\emph{FGCS}} (\bibinfo{year}{2020}),
  \bibinfo{pages}{320--333}.
\newblock


\bibitem[Shirahama et~al\mbox{.}(2004)]%
        {shirahama2004video}
\bibfield{author}{\bibinfo{person}{Kimiaki Shirahama},
  \bibinfo{person}{Kazuhisa Iwamoto}, {and} \bibinfo{person}{Kuniaki Uehera}.}
  \bibinfo{year}{2004}\natexlab{}.
\newblock \showarticletitle{Video data mining: rhythms in a movie}. In
  \bibinfo{booktitle}{\emph{ICME}}. \bibinfo{pages}{1463--1466}.
\newblock


\bibitem[Su et~al\mbox{.}(2022)]%
        {su2022cross}
\bibfield{author}{\bibinfo{person}{Hongzu Su}, \bibinfo{person}{Yifei Zhang},
  \bibinfo{person}{Xuejiao Yang}, \bibinfo{person}{Hua Hua},
  \bibinfo{person}{Shuangyang Wang}, {and} \bibinfo{person}{Jingjing Li}.}
  \bibinfo{year}{2022}\natexlab{}.
\newblock \showarticletitle{Cross-domain Recommendation via Adversarial
  Adaptation}. In \bibinfo{booktitle}{\emph{CIKM}}.
  \bibinfo{pages}{1808--1817}.
\newblock


\bibitem[Sun et~al\mbox{.}(2019)]%
        {sun2019multi}
\bibfield{author}{\bibinfo{person}{Jianing Sun}, \bibinfo{person}{Yingxue
  Zhang}, \bibinfo{person}{Chen Ma}, \bibinfo{person}{Mark Coates},
  \bibinfo{person}{Huifeng Guo}, \bibinfo{person}{Ruiming Tang}, {and}
  \bibinfo{person}{Xiuqiang He}.} \bibinfo{year}{2019}\natexlab{}.
\newblock \showarticletitle{Multi-Graph Convolution Collaborative Filtering}.
  In \bibinfo{booktitle}{\emph{ICDM}}. \bibinfo{pages}{1306--1311}.
\newblock


\bibitem[Wang et~al\mbox{.}(2022c)]%
        {wang2022inter}
\bibfield{author}{\bibinfo{person}{Ke Wang}, \bibinfo{person}{Yanmin Zhu},
  \bibinfo{person}{Haobing Liu}, \bibinfo{person}{Tianzi Zang},
  \bibinfo{person}{Chunyang Wang}, {and} \bibinfo{person}{Kuan Liu}.}
  \bibinfo{year}{2022}\natexlab{c}.
\newblock \showarticletitle{Inter-and Intra-Domain Relation-Aware Heterogeneous
  Graph Convolutional Networks for Cross-Domain Recommendation}. In
  \bibinfo{booktitle}{\emph{DASFAA}}. \bibinfo{pages}{53--68}.
\newblock


\bibitem[Wang et~al\mbox{.}(2022a)]%
        {wang2022disentangledsurvey}
\bibfield{author}{\bibinfo{person}{Xin Wang}, \bibinfo{person}{Hong Chen},
  \bibinfo{person}{Si'ao Tang}, \bibinfo{person}{Zihao Wu}, {and}
  \bibinfo{person}{Wenwu Zhu}.} \bibinfo{year}{2022}\natexlab{a}.
\newblock \showarticletitle{Disentangled Representation Learning}.
\newblock \bibinfo{journal}{\emph{arXiv preprint arXiv:2211.11695}}
  (\bibinfo{year}{2022}).
\newblock


\bibitem[Wang et~al\mbox{.}(2022b)]%
        {wang2022disentangled}
\bibfield{author}{\bibinfo{person}{Xin Wang}, \bibinfo{person}{Hong Chen},
  \bibinfo{person}{Yuwei Zhou}, \bibinfo{person}{Jianxin Ma}, {and}
  \bibinfo{person}{Wenwu Zhu}.} \bibinfo{year}{2022}\natexlab{b}.
\newblock \showarticletitle{Disentangled Representation Learning for
  Recommendation}.
\newblock \bibinfo{journal}{\emph{TPAMI}} (\bibinfo{year}{2022}),
  \bibinfo{pages}{408--424}.
\newblock


\bibitem[Wang et~al\mbox{.}(2019a)]%
        {wang2019kgat}
\bibfield{author}{\bibinfo{person}{Xiang Wang}, \bibinfo{person}{Xiangnan He},
  \bibinfo{person}{Yixin Cao}, \bibinfo{person}{Meng Liu}, {and}
  \bibinfo{person}{Tat-Seng Chua}.} \bibinfo{year}{2019}\natexlab{a}.
\newblock \showarticletitle{KGAT: Knowledge Graph Attention Network for
  Recommendation}. In \bibinfo{booktitle}{\emph{KDD}}.
  \bibinfo{pages}{950--958}.
\newblock


\bibitem[Wang et~al\mbox{.}(2019b)]%
        {wang2019neural}
\bibfield{author}{\bibinfo{person}{Xiang Wang}, \bibinfo{person}{Xiangnan He},
  \bibinfo{person}{Meng Wang}, \bibinfo{person}{Fuli Feng}, {and}
  \bibinfo{person}{Tat-Seng Chua}.} \bibinfo{year}{2019}\natexlab{b}.
\newblock \showarticletitle{Neural Graph Collaborative Filtering}. In
  \bibinfo{booktitle}{\emph{SIGIR}}. \bibinfo{pages}{165--174}.
\newblock


\bibitem[Wang et~al\mbox{.}(2020)]%
        {wang2020disentangled}
\bibfield{author}{\bibinfo{person}{Xiang Wang}, \bibinfo{person}{Hongye Jin},
  \bibinfo{person}{An Zhang}, \bibinfo{person}{Xiangnan He},
  \bibinfo{person}{Tong Xu}, {and} \bibinfo{person}{Tat-Seng Chua}.}
  \bibinfo{year}{2020}\natexlab{}.
\newblock \showarticletitle{Disentangled Graph Collaborative Filtering}. In
  \bibinfo{booktitle}{\emph{SIGIR}}. \bibinfo{pages}{1001--1010}.
\newblock


\bibitem[Wang et~al\mbox{.}(2021)]%
        {wang2021mixup}
\bibfield{author}{\bibinfo{person}{Yiwei Wang}, \bibinfo{person}{Wei Wang},
  \bibinfo{person}{Yuxuan Liang}, \bibinfo{person}{Yujun Cai}, {and}
  \bibinfo{person}{Bryan Hooi}.} \bibinfo{year}{2021}\natexlab{}.
\newblock \showarticletitle{Mixup for Node and Graph Classification}. In
  \bibinfo{booktitle}{\emph{WWW}}. \bibinfo{pages}{3663--3674}.
\newblock


\bibitem[Xiao et~al\mbox{.}(2023)]%
        {xiao2023catcl}
\bibfield{author}{\bibinfo{person}{Shuo Xiao}, \bibinfo{person}{Dongqing Zhu},
  \bibinfo{person}{Chaogang Tang}, {and} \bibinfo{person}{Zhenzhen Huang}.}
  \bibinfo{year}{2023}\natexlab{}.
\newblock \showarticletitle{CATCL: Joint Cross-Attention Transfer and
  Contrastive Learning for Cross-Domain Recommendation}. In
  \bibinfo{booktitle}{\emph{DASFAA}}. \bibinfo{pages}{446--461}.
\newblock


\bibitem[Yuan et~al\mbox{.}(2019)]%
        {ijcai2019p587}
\bibfield{author}{\bibinfo{person}{Feng Yuan}, \bibinfo{person}{Lina Yao},
  {and} \bibinfo{person}{Boualem Benatallah}.} \bibinfo{year}{2019}\natexlab{}.
\newblock \showarticletitle{DARec: Deep Domain Adaptation for Cross-Domain
  Recommendation via Transferring Rating Patterns}. In
  \bibinfo{booktitle}{\emph{IJCAI}}. \bibinfo{pages}{4227--4233}.
\newblock


\bibitem[Zang et~al\mbox{.}(2022)]%
        {zang2021survey}
\bibfield{author}{\bibinfo{person}{Tianzi Zang}, \bibinfo{person}{Yanmin Zhu},
  \bibinfo{person}{Haobing Liu}, \bibinfo{person}{Ruohan Zhang}, {and}
  \bibinfo{person}{Jiadi Yu}.} \bibinfo{year}{2022}\natexlab{}.
\newblock \showarticletitle{A Survey on Cross-Domain Recommendation:
  Taxonomies, Methods, and Future Directions}.
\newblock \bibinfo{journal}{\emph{TOIS}} (\bibinfo{year}{2022}),
  \bibinfo{pages}{1--39}.
\newblock


\bibitem[Zhang et~al\mbox{.}(2018)]%
        {zhang2017mixup}
\bibfield{author}{\bibinfo{person}{Hongyi Zhang}, \bibinfo{person}{Moustapha
  Ciss{\'{e}}}, \bibinfo{person}{Yann~N. Dauphin}, {and} \bibinfo{person}{David
  Lopez{-}Paz}.} \bibinfo{year}{2018}\natexlab{}.
\newblock \showarticletitle{\emph{mixup}: Beyond Empirical Risk Minimization}.
  In \bibinfo{booktitle}{\emph{ICLR}}.
\newblock


\bibitem[Zhang et~al\mbox{.}(2023)]%
        {zhang2023disentangled}
\bibfield{author}{\bibinfo{person}{Ruohan Zhang}, \bibinfo{person}{Tianzi
  Zang}, \bibinfo{person}{Yanmin Zhu}, \bibinfo{person}{Chunyang Wang},
  \bibinfo{person}{Ke Wang}, {and} \bibinfo{person}{Jiadi Yu}.}
  \bibinfo{year}{2023}\natexlab{}.
\newblock \showarticletitle{Disentangled Contrastive Learning for Cross-Domain
  Recommendation}. In \bibinfo{booktitle}{\emph{DASFAA}}.
  \bibinfo{pages}{163--178}.
\newblock


\bibitem[Zhang et~al\mbox{.}(2022)]%
        {zhang2022multi}
\bibfield{author}{\bibinfo{person}{Xinyue Zhang}, \bibinfo{person}{Jingjing
  Li}, \bibinfo{person}{Hongzu Su}, \bibinfo{person}{Lei Zhu}, {and}
  \bibinfo{person}{Heng~Tao Shen}.} \bibinfo{year}{2022}\natexlab{}.
\newblock \showarticletitle{Multi-Level Attention-Based Domain Disentanglement
  for Bidirectional Cross-Domain Recommendation}.
\newblock \bibinfo{journal}{\emph{TOIS}} (\bibinfo{year}{2022}).
\newblock


\bibitem[Zhang et~al\mbox{.}(2020)]%
        {zhang2020content}
\bibfield{author}{\bibinfo{person}{Yin Zhang}, \bibinfo{person}{Ziwei Zhu},
  \bibinfo{person}{Yun He}, {and} \bibinfo{person}{James Caverlee}.}
  \bibinfo{year}{2020}\natexlab{}.
\newblock \showarticletitle{Content-Collaborative Disentanglement
  Representation Learning for Enhanced Recommendation}. In
  \bibinfo{booktitle}{\emph{RecSys}}. \bibinfo{pages}{43--52}.
\newblock


\bibitem[Zhao et~al\mbox{.}(2019)]%
        {zhao2019cross}
\bibfield{author}{\bibinfo{person}{Cheng Zhao}, \bibinfo{person}{Chenliang Li},
  {and} \bibinfo{person}{Cong Fu}.} \bibinfo{year}{2019}\natexlab{}.
\newblock \showarticletitle{Cross-Domain Recommendation via Preference
  Propagation Graphnet}. In \bibinfo{booktitle}{\emph{CIKM}}.
  \bibinfo{pages}{2165--2168}.
\newblock


\bibitem[Zhao et~al\mbox{.}(2022)]%
        {zhao2022multi}
\bibfield{author}{\bibinfo{person}{Xiaoyun Zhao}, \bibinfo{person}{Ning Yang},
  {and} \bibinfo{person}{Philip~S Yu}.} \bibinfo{year}{2022}\natexlab{}.
\newblock \showarticletitle{Multi-Sparse-Domain Collaborative Recommendation
  via Enhanced Comprehensive Aspect Preference Learning}. In
  \bibinfo{booktitle}{\emph{WSDM}}. \bibinfo{pages}{1452--1460}.
\newblock


\bibitem[Zhu et~al\mbox{.}(2019)]%
        {zhu2019dtcdr}
\bibfield{author}{\bibinfo{person}{Feng Zhu}, \bibinfo{person}{Chaochao Chen},
  \bibinfo{person}{Yan Wang}, \bibinfo{person}{Guanfeng Liu}, {and}
  \bibinfo{person}{Xiaolin Zheng}.} \bibinfo{year}{2019}\natexlab{}.
\newblock \showarticletitle{DTCDR: A Framework for Dual-Target Cross-Domain
  Recommendation}. In \bibinfo{booktitle}{\emph{CIKM}}.
  \bibinfo{pages}{1533–1542}.
\newblock


\bibitem[Zhu et~al\mbox{.}(2018)]%
        {ijcai2018p516}
\bibfield{author}{\bibinfo{person}{Feng Zhu}, \bibinfo{person}{Yan Wang},
  \bibinfo{person}{Chaochao Chen}, \bibinfo{person}{Guanfeng Liu},
  \bibinfo{person}{Mehmet Orgun}, {and} \bibinfo{person}{Jia Wu}.}
  \bibinfo{year}{2018}\natexlab{}.
\newblock \showarticletitle{A Deep Framework for Cross-Domain and Cross-System
  Recommendations}. In \bibinfo{booktitle}{\emph{IJCAI}}.
  \bibinfo{pages}{3711--3717}.
\newblock


\bibitem[Zhu et~al\mbox{.}(2020)]%
        {zhu2020graphical}
\bibfield{author}{\bibinfo{person}{Feng Zhu}, \bibinfo{person}{Yan Wang},
  \bibinfo{person}{Chaochao Chen}, \bibinfo{person}{Guanfeng Liu}, {and}
  \bibinfo{person}{Xiaolin Zheng}.} \bibinfo{year}{2020}\natexlab{}.
\newblock \showarticletitle{A Graphical and Attentional Framework for
  Dual-Target Cross-Domain Recommendation}. In
  \bibinfo{booktitle}{\emph{IJCAI}}. \bibinfo{pages}{3001--3008}.
\newblock


\bibitem[Zhu et~al\mbox{.}(2021a)]%
        {ijcai2021p639}
\bibfield{author}{\bibinfo{person}{Feng Zhu}, \bibinfo{person}{Yan Wang},
  \bibinfo{person}{Chaochao Chen}, \bibinfo{person}{Jun Zhou},
  \bibinfo{person}{Longfei Li}, {and} \bibinfo{person}{Guanfeng Liu}.}
  \bibinfo{year}{2021}\natexlab{a}.
\newblock \showarticletitle{Cross-Domain Recommendation: Challenges, Progress,
  and Prospects}. In \bibinfo{booktitle}{\emph{IJCAI}}.
  \bibinfo{pages}{4721--4728}.
\newblock


\bibitem[Zhu et~al\mbox{.}(2021b)]%
        {zhu2021unified}
\bibfield{author}{\bibinfo{person}{Feng Zhu}, \bibinfo{person}{Yan Wang},
  \bibinfo{person}{Jun Zhou}, \bibinfo{person}{Chaochao Chen},
  \bibinfo{person}{Longfei Li}, {and} \bibinfo{person}{Guanfeng Liu}.}
  \bibinfo{year}{2021}\natexlab{b}.
\newblock \showarticletitle{A Unified Framework for Cross-Domain and
  Cross-System Recommendations}.
\newblock \bibinfo{journal}{\emph{TKDE}} (\bibinfo{year}{2021}),
  \bibinfo{pages}{1171--1184}.
\newblock


\bibitem[Zhu et~al\mbox{.}(2021c)]%
        {zhu2021learning}
\bibfield{author}{\bibinfo{person}{Yongchun Zhu}, \bibinfo{person}{Ruobing
  Xie}, \bibinfo{person}{Fuzhen Zhuang}, \bibinfo{person}{Kaikai Ge},
  \bibinfo{person}{Ying Sun}, \bibinfo{person}{Xu Zhang}, \bibinfo{person}{Leyu
  Lin}, {and} \bibinfo{person}{Juan Cao}.} \bibinfo{year}{2021}\natexlab{c}.
\newblock \showarticletitle{Learning to Warm Up Cold Item Embeddings for
  Cold-start Recommendation with Meta Scaling and Shifting Networks}. In
  \bibinfo{booktitle}{\emph{SIGIR}}. \bibinfo{pages}{1167--1176}.
\newblock


\end{thebibliography}

\end{document}